\documentclass[reprint,amsmath,amssymb,aps,prd,noeprint,nolongbibliography,superscriptaddress,nofootinbib]{revtex4-2}
\usepackage{graphicx,xcolor,verbatim,makecell,multirow,float,mathtools,slashed,booktabs,soul,makecell} 
\usepackage[colorlinks,citecolor=blue,urlcolor=blue,linkcolor=blue]{hyperref}
\usepackage[normalem]{ulem}
\makeatletter
\def\UTFviii@defined#1{\ifx#1\relax!!FIXME!!\else\expandafter#1\fi} 
\makeatother
\begin{document}
\title{LHC Mono-\texorpdfstring{$W/Z$}{WZ} Signatures as a Probe for Dark Matter Explanations of Astrophysical Excesses}
\author{Yu-Chen Guo}
\email{ycguo@lnnu.edu.cn}
\affiliation{Center for Theoretical and Experimental High Energy Physics, Department of Physics,
Liaoning Normal University, Dalian 116029, China}
\author{Ying-Xin Li}
\affiliation{Center for Theoretical and Experimental High Energy Physics, Department of Physics,
Liaoning Normal University, Dalian 116029, China}
\author{Chih-Ting Lu}
\email{ctlu@njnu.edu.cn}
\affiliation{Department of Physics and Institute of Theoretical Physics, \\ Nanjing Normal University, Nanjing, 210023, China}
\affiliation{Nanjing Key Laboratory of Particle Physics and Astrophysics, Nanjing, 210023, China}
{\hfill CPTNP-2026-015}
\begin{abstract}
The inert two-Higgs doublet model (IDM) is a compelling framework for weakly interacting massive particles (WIMPs) linked to electroweak symmetry breaking. It can account for both the Galactic Center gamma‑ray excess (GCE) and the AMS‑02 antiproton anomaly while also satisfying relic density and direct detection constraints for dark matter (DM) masses in the $55–75$ GeV range. Three specific DM annihilation channels can be identified: Higgs resonance, $SA$ co‑annihilation, and $SS\to WW^{\ast}$ annihilation. Among these, the DM mass range of $70-75$ GeV with dominant $SS\to WW^{\ast}$ annihilation has received less attention in collider searches. To validate this parameter space, we combine LHC searches for mono‑$W/Z$ signatures. In particular, we develop a channel‑separation strategy to disentangle the contributions of charged mass splitting ($\Delta^{\pm}$) and neutral mass splitting ($\Delta^0$) in the inert scalar sector at the LHC. Our results indicate that most of the parameter space consistent with these astrophysical anomalies in the $SS\to WW^{\ast}$ annihilation regime will be testable at the High‑Luminosity LHC. Specifically, from the leptonic channel we obtain a $2\sigma$ exclusion limit of $80 \lesssim \Delta^0 \lesssim 260$ GeV, while the hadronic channel yields $30 \lesssim \Delta^0 \lesssim 150$ GeV and $70 \lesssim \Delta^{\pm} \lesssim 230$ GeV for $m_S = 70$ GeV. 
\end{abstract}
\maketitle

\flushbottom

\section{Introduction}

Dark matter (DM) constitutes approximately $27\%$ of the total energy density of the Universe, yet its fundamental particle nature remains one of the most profound mysteries in particle physics, astrophysics, and cosmology~\cite{Trimble:1987ee,Bertone:2004pz,Arbey:2021gdg,Cirelli:2024ssz}. Among various indirect detection signals, two particularly intriguing anomalies are the Galactic Center gamma-ray excess (GCE)~\cite{Daylan:2014rsa,Slatyer:2021qgc,Cholis:2021rpp,DiMauro:2021qcf} and the AMS-02 antiproton anomaly~\cite{Cui:2016ppb,Cuoco:2016eej,Calore:2022stf}. The GCE manifests as an unexpected excess of gamma rays in the $1-3$ GeV energy range originating from the Galactic Center, while AMS-02 has reported an antiproton flux in the $10-20$ GeV range that significantly exceeds conventional astrophysical background predictions. Despite extensive scrutiny, the interpretation of these anomalies remains a subject of intense debate due to uncertainties related to unresolved populations of millisecond pulsars and cosmic-ray propagation models~\cite{Bartels:2015aea,Heisig:2020nse,Gautam:2021wqn,Lv:2023gdt,Duan:2025ead}.

In this context, the inert two-Higgs doublet model (IDM)~\cite{Deshpande:1977rw,Barbieri:2006dq,LopezHonorez:2006gr} emerges as a highly compelling theoretical framework. The IDM extends the Standard Model (SM) by introducing an additional Higgs doublet, $H_D$, which is rendered ``inert'' (i.e., it does not acquire a vacuum expectation value) by an exact $Z_2$ symmetry. This construction naturally furnishes a stable DM candidate in the form of the lightest neutral component of $H_D$. Recent global analyses have demonstrated that, for DM masses situated in the $55-75$ GeV window, the IDM can simultaneously accommodate both the GCE and the AMS-02 antiproton anomaly~\cite{Fan:2022wmass,Fan:2024wvo}\footnote{Other DM models can also explain either the GCE~\cite{Dolan:2014ska,Ipek:2014gua,Berlin:2014tja,Alves:2014yha,Agrawal:2014una,Izaguirre:2014vva,Ko:2014gha,Abdullah:2014lla,Martin:2014sxa,Berlin:2014pya,Cheung:2014lqa,Agrawal:2014oha} or the AMS-02 antiproton anomaly~\cite{Clark:2017fum,Jia:2017kjw,Cui:2018nlm}, and in some cases even both~\cite{Arcadi:2017wis,Abdughani:2021pdc,DiMauro:2023tho,Fan:2024wvo}.}. Within this mass range, DM annihilation processes, such as the Higgs resonance and the gauge-dominated $SS \to WW^\ast$ channels, can seamlessly yield the required thermal relic abundance and astrophysical fluxes.

However, thoroughly probing this parameter space requires a multifaceted strategy. Stringent limits from direct detection experiments, such as PandaX-4T~\cite{PandaX:2024qfu}, LZ~\cite{LZ:2024zvo}, XENONnT~\cite{XENON:2025vwd}, have already severely constrained the tree-level Higgs portal coupling. This necessitates the inclusion of loop-level corrections to accurately predict the nuclear recoil cross-sections~\cite{Abe:2015rja,Tsai:2019eqi}, revealing that the $SS \to WW^\ast$ channel can remain viable even when the Higgs portal coupling is vanishingly small. Furthermore, definitively disentangling a DM signal from standard astrophysical backgrounds requires cross-validation from high-energy colliders. 

To address this, we propose a novel and complementary search strategy at the Large Hadron Collider (LHC) focusing on mono-$W$ and mono-$Z$ signatures. By leveraging the unique kinematic topologies of the IDM in the low-mass window, our analysis introduces a channel-separation technique: the mono-$W$ signature ($pp \to h^\pm S \to W^\pm S S$) serves as a direct probe for the charged mass splitting ($\Delta^\pm$), whereas the mono-$Z$ signature ($pp \to A S \to Z S S$) provides independent constraints on the neutral mass splitting ($\Delta^0$). Although the invisible channel of vector‑boson fusion can constrain the relevant parameter space (as shown in Refs.~\cite{Datta:2016fns,Belyaev:2016lok,Dercks:2018wch,Kalinowski:2020rmb}), this approach cannot disentangle the contributions of $\Delta^\pm$, $\Delta^0$, and other model parameters. Therefore, our search strategy differs from previous studies while offering complementary results\footnote{Other proposals for IDM searches at the LHC can be found in Refs.~\cite{Cao:2007rm,Belanger:2015kga,Ilnicka:2015jba,Datta:2016nfz,Belyaev:2016lok,Dercks:2018wch,Belyaev:2018ext,Kalinowski:2020rmb}; however, their analysis strategies and/or motivations differ from those of this work. Moreover, preliminary results from the CMS Collaboration have appeared in Ref.~\cite{CMS:2026rhc}.}. 

The remainder of this paper is organized as follows. In Section~\ref{sec:model}, we provide a brief review of the theoretical framework of the IDM. Section~\ref{sec:analysis} details our parameter selection rationale and signal topologies. In Section~\ref{sec:result}, we present our rigorous LHC simulation strategies for both the leptonic and hadronic channels, including background analyses and kinematic selections. Finally, Section~\ref{sec:conclusion} concludes with a summary of our key findings. 

\section{A Brief Review of the Inert Two Higgs Doublet Model} 
\label{sec:model}

The IDM represents one of the most minimal extensions of the SM capable of accommodating a viable DM candidate. Originally proposed by Deshpande and Ma~\cite{Deshpande:1977rw}, it extends the SM scalar sector by introducing a second Higgs doublet, $H_D$, which is odd under an imposed, exact $Z_2$ symmetry, whereas all SM particles are $Z_2$-even.

\subsection{Model Framework and Scalar Potential} 

The two scalar doublets are parameterized as:
\begin{equation}
H = \begin{pmatrix} G^+ \\ \frac{1}{\sqrt{2}}(v + h + iG^0) \end{pmatrix}, \quad H_D = \begin{pmatrix} h^+ \\ \frac{1}{\sqrt{2}}(S + iA) \end{pmatrix},
\end{equation}
where \(G^\pm\) and \(G^0\) denote the Goldstone bosons that are absorbed by the \(W^\pm\) and \(Z\) bosons during electroweak symmetry breaking (EWSB), and \(v \approx 246\) GeV is the SM Higgs vacuum expectation value (VEV). 
The exact $Z_2$ symmetry prevents $H_D$ from acquiring a VEV and forbids any direct Yukawa couplings to SM fermions, ensuring the absolute stability of the lightest $Z_2$-odd state ($S$ or $A$). 
The general CP-conserving scalar potential is given by:
\begin{align}
V = &\mu_1^2 |H|^2 + \mu_2^2 |H_D|^2 + \lambda_1 |H|^4 + \lambda_2 |H_D|^4 + \lambda_3 |H|^2 |H_D|^2 \nonumber\\
&+ \lambda_4 |H^\dagger H_D|^2 + \frac{\lambda_5}{2}\left[(H^\dagger H_D)^2 + \text{h.c.}\right].
\end{align}
Here $\mu_i$ ($i=1,2$) has the same dimension as the mass, while $\lambda_i$ ($i=1,\dots,5$) is dimensionless. After EWSB, the physical spectrum of the inert sector comprises four scalar particles: a CP-even scalar $S$, a CP-odd pseudoscalar $A$, and a pair of charged scalars $h^\pm$. In this work, we assume that the CP-even scalar $S$ is the lightest $Z_2$-odd particle (LOP) and thus the DM candidate.

The physical masses of the scalar particles can be expressed in terms of the potential parameters. For instance, one obtains:
\begin{align}
m_h^2 &= 2\lambda_1 v^2, \\
m_S^2 &= \mu_2^2 + \frac{1}{2}(\lambda_3 + \lambda_4 + \lambda_5)v^2 \equiv \mu_2^2 + \lambda_S v^2, \\
m_A^2 &= \mu_2^2 + \frac{1}{2}(\lambda_3 + \lambda_4 - \lambda_5)v^2, \\
m_{h^\pm}^2 &= \mu_2^2 + \frac{1}{2}\lambda_3 v^2.
\end{align}
The mass splittings within the dark sector, defined as $\Delta^0 = m_A - m_S$ and $\Delta^\pm = m_{h^\pm} - m_S$, predominantly dictate the kinematic topologies of the collider signatures. 
The effective Higgs-DM coupling $\lambda_S$ plays a critical role in determining the thermal relic density and direct detection rates.

\subsection{Theoretical Constraints and Phenomenological Implications}

The IDM is subject to a set of rigorous theoretical constraints that ensure the vacuum stability and perturbative consistency of the scalar potential. To maintain perturbativity, all quartic couplings are generally required to satisfy $|\lambda_i| \lesssim 8\pi$. Furthermore, the requirement that the scalar potential is bounded from below imposes the following strict conditions:
\begin{equation}
\lambda_1, \lambda_2 > 0, \quad \lambda_3 + 2\sqrt{\lambda_1 \lambda_2} > 0, \quad \lambda_3 + \lambda_4 - |\lambda_5| + 2\sqrt{\lambda_1 \lambda_2} > 0.
\end{equation}
Additionally, tree-level perturbative unitarity constraints derived from various scalar and gauge boson $2 \to 2$ scattering processes further restrict the viable parameter space. These theoretical conditions are taken into account in our study using 2HDMC~\cite{Eriksson:2009ws}.

Moreover, the inert nature of $H_D$ guaranties the absence of tree-level flavor-changing neutral currents (FCNCs) and prevents  additional scalars from directly coupling to SM fermions. This elegant feature, combined with the absolute stability of the LOP enforced by the exact $Z_2$ symmetry, renders the IDM a highly predictive and minimal framework for scalar DM.

From a phenomenological perspective, the IDM is particularly compelling because the dark sector interacts with the SM not only via the Higgs portal but also strongly through the $W$ and $Z$ gauge bosons. The coupling between the DM candidate $S$ and the SM Higgs boson is governed by $\lambda_S$, which critically determines both the thermal relic density and the DM-nucleon elastic scattering cross-section at the tree-level. In the low-mass DM regime, the three dominant DM annihilation channels are the Higgs resonance, co‑annihilation, and $WW^{\ast}$ annihilation. All of them require a small $\lambda_S$ to reproduce the observed relic abundance~\cite{Tsai:2019eqi}. The heavily suppressed $\lambda_S$ also substantially reduces the tree‑level DM‑nucleon scattering rate. This behavior intrinsically necessitates the inclusion of loop‑level corrections, which become the dominant contribution to direct detection cross sections in the small Higgs portal limit~\cite{Abe:2015rja,Tsai:2019eqi}. However, there remains a significant region of parameter space below the predicted neutrino floor regime where direct detection experiments largely lose their ability to distinguish between DM‑induced events and neutrino backgrounds. In this regime, collider searches become an important complementary way for exploring the viable parameter space. 

Therefore, to highlight the gauge interaction dominated regime (i.e., $SS \to WW^{\ast}$) that evades stringent direct detection limits, we establish a benchmark scenario with a highly suppressed Higgs portal coupling ($\lambda_S \lesssim 10^{-3}$). Based on this setup, current direct detection constraints~\cite{PandaX:2024qfu,LZ:2024zvo,XENON:2025vwd} are evaded. Such a tiny $\lambda_S$ value is very challenging to probe via invisible Higgs decays even at the HL‑LHC~\cite{deBlas:2019rxi,CMS:2022dwd}. Moreover, as we will show in the next section, the values of $\lambda_S$ allowed by current constraints have little impact on our collider studies. Additionally, in the case of compressed mass spectra among $S$, $A$, and $h^{\pm}$ that give rise to significant co‑annihilation effects, current and future soft‑lepton search strategies are expected to cover the relevant viable parameter space~\cite{Tsai:2019eqi,Fan:2022wmass,Fan:2024wvo}. Hence, exploring the key parameter space corresponding to $SS \to WW^{\ast}$ is the main focus of this work. 

\section{Mono-\texorpdfstring{$W/Z$}{WZ} Signatures at the LHC}
\label{sec:analysis}

The primary objective of our collider analysis is to investigate the production of DM pairs in association with a SM electroweak gauge boson $(V=W,Z)$, leading to the distinctive mono-$V$ plus missing transverse momentum signatures. In the IDM, these processes provide a direct probe of the dark sector's gauge interactions and its mass spectrum. Before presenting the details of our Monte Carlo simulation, we first establish the phenomenological motivations that dictate our choice of the benchmark parameter space. 

\subsection{Phenomenological Motivations and Benchmark Selection}

Recent global fits and phenomenological analyses~\cite{Zhu:2022excess, Fan:2022wmass} have highlighted the IDM as a robust candidate to explain two persistent astrophysical anomalies: the Galactic Center gamma-ray excess (GCE) observed by Fermi-LAT~\cite{Fermi-LAT:2015sau} and the antiproton excess reported by AMS-02~\cite{AMS:2016oqu}. To simultaneously accommodate the observed relic density via thermal freeze-out and provide the required annihilation cross-section for these indirect detection signals, the DM mass is typically constrained within the low-mass DM window of $54 \lesssim m_S \lesssim 74$~GeV. 

\begin{table}[htbp]
\caption{Benchmark parameter ranges of the IDM favored by the Fermi-LAT GCE and AMS-02 antiproton excess.}  
\label{tab:para}
\centering
\renewcommand{\arraystretch}{1.3}
\begin{tabular}{cc}
\toprule  
IDM Parameter & Benchmark Range \\  
\midrule  
 $\lambda_S$    & $[-2.417 \times 10^{-3}, \, 2.921 \times 10^{-3}]$ \\
 $m_{S}$        & $[70~\mathrm{GeV}, \, 75~\mathrm{GeV}]$  \\
 $m_{A}$        & $[70.42~\mathrm{GeV}, \, 556.10~\mathrm{GeV}]$ \\
 $m_{h^{\pm}}$  & $[140.83~\mathrm{GeV}, \, 571.56~\mathrm{GeV}]$ \\
\bottomrule  
\end{tabular}
\end{table}

Guided by these phenomenological constraints, we establish the viable parameter space for our numerical analysis, as summarized in Table~\ref{tab:para}.\footnote{The coupling $\lambda_2$ governs the dark sector self‑interaction and only affects the DM self‑interaction and loop corrections, which are not relevant to our collider searches. In our numerical studies, we set $\lambda_2 = 5\times 10^{-4}$.} Specifically, the parameter regime favoring the $SS \to WW^\ast$ annihilation channel ($m_S \simeq 70\text{--}74$~GeV) exhibits the highest degree of compatibility with the spectral shapes of both astrophysical excesses. In light of these astrophysical cues, we focus our collider study on the benchmark mass window of $m_S \in [70, 75]$~GeV. Within this range, the DM candidate $S$ is sufficiently heavy to evade the stringent bounds from invisible Higgs decays ($h \to SS$), yet light enough to be copiously produced at the LHC. 

\begin{figure*}[htbp]
\centering
\includegraphics[width=0.8\textwidth]{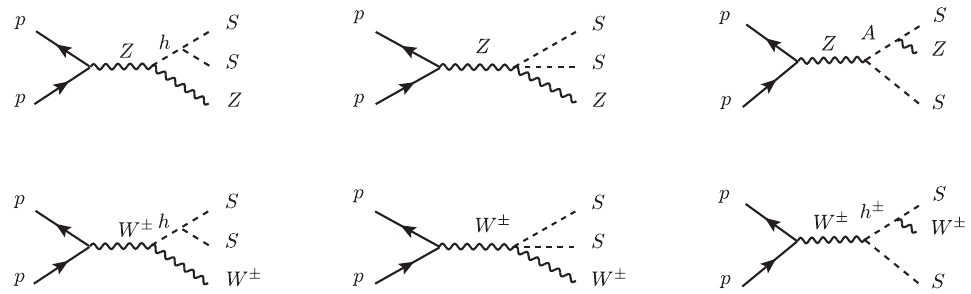}
\caption{Representative Feynman diagrams for the signal process $pp \to ZSS$ (top) and $pp \to W^{\pm}SS$ (bottom).}
\label{fig:feynman}
\end{figure*}

At the LHC, the production of $SS$ pairs in association with a $Z$ boson ($pp \to ZSS$) is mediated primarily by the exchange of the pseudoscalar $A$. Conversely, the mono-$W$ signature ($pp \to W^\pm SS$) is governed by the exchange of charged scalars $h^\pm$. The representative Feynman diagrams for these signal processes are displayed in Figure~\ref{fig:feynman}.

\subsection{Signal Cross-section Dependence on Model Parameters}

\begin{figure}[htbp]
\centering
\includegraphics[width=0.45\textwidth]{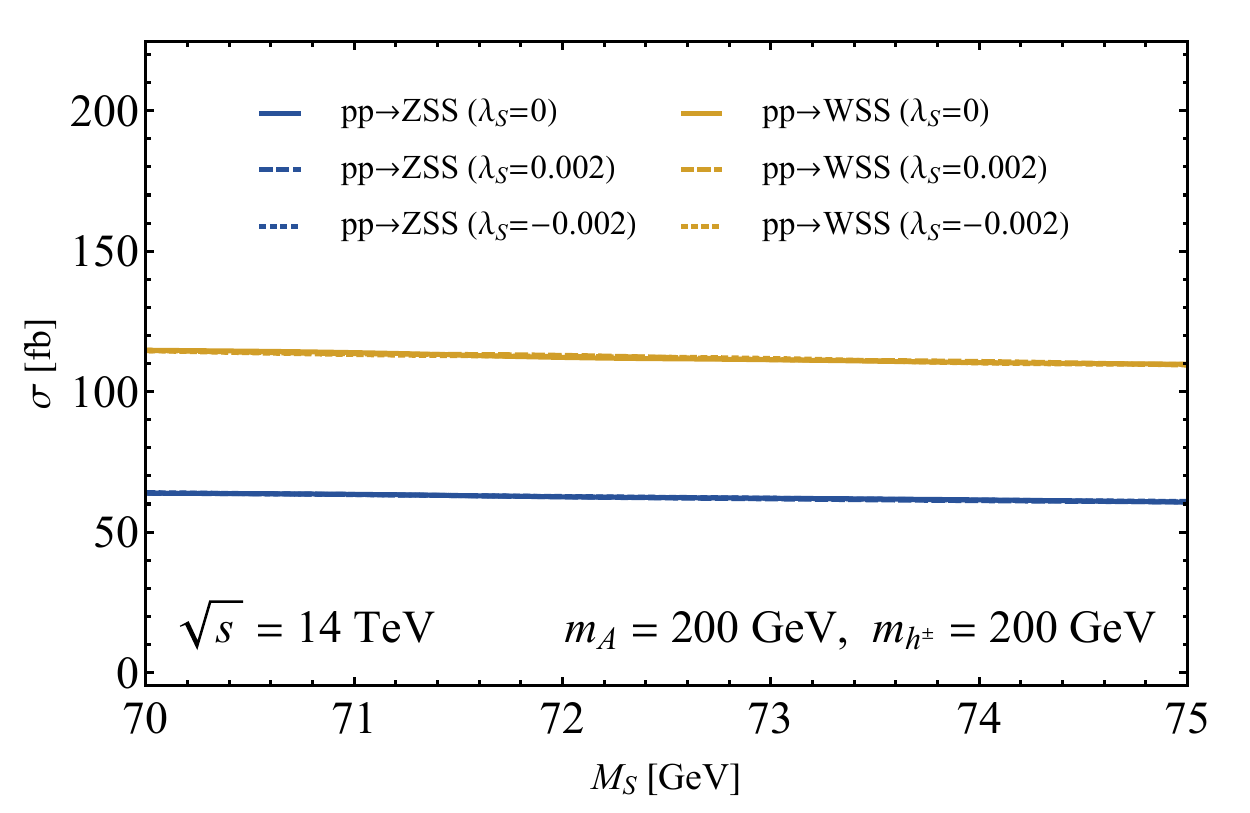}
\caption{Production cross-sections for $pp \to ZSS$ and $pp \to W^{\pm}SS$ as functions of $m_S$ at $\sqrt{s}=14$~TeV. The near-overlap of curves for different $\lambda_S$ values highlights the dominance of gauge interactions.}
\label{fig:Ms-pp_vss}
\end{figure}

To understand the sensitivity of the signal strength to the underlying model parameters, we evaluated the production cross-sections as functions of the DM mass $m_S$ and the Higgs-portal coupling $\lambda_S$. The UFO model file for the IDM is generated using \texttt{FeynRules}~\cite{feynrules}, and then the Monte Carlo event generator \texttt{MadGraph5\_aMC@NLO}~\cite{madgraph} is used to numerically calculate the signal production cross sections. In Figure~\ref{fig:Ms-pp_vss}, we present the cross-sections for both mono-$Z$ and mono-$W$ processes at $\sqrt{s}=14$~TeV, assuming fixed inert scalar masses $m_A = m_{h^\pm} = 200$~GeV. 

Several key features emerge from this numerical result. First, the cross-sections for both $ZSS$ and $W^\pm SS$ production exhibit a remarkable insensitivity to the variation of $m_S$ within the target $[70, 75]$~GeV interval. Second, the Higgs-portal coupling $\lambda_S$ is already subject to severe constraints from global fits, with $|\lambda_S| \lesssim 2\times 10^{-2}$~\cite{Fan:2022wmass,Fan:2024wvo}. We further find that the curves corresponding to different values of $\lambda_S$ within this allowed range are nearly indistinguishable and overlap almost perfectly. This behavior indicates an effective decoupling of the signal production from the Higgs sector in this mass regime. The signal strength is predominantly dictated by the electroweak gauge couplings and the kinematic phase space determined by the inert scalar mass spectrum, rather than the Higgs-portal interaction. Consequently, to ensure that our results are conservative and focus purely on gauge-mediated contributions, we adopt a minimal Higgs-portal coupling $\lambda_S = 10^{-3}$ as our benchmark value for subsequent analysis.

The interplay between the masses of the inert scalars plays a decisive role in the collider phenomenology. The kinematic thresholds for on-shell resonance production serve as the most significant factor in enhancing the signal visibility. For the mono-$Z$ channel, the intermediate pseudoscalar $A$ can undergo a two-body on-shell decay ($A \to ZS$) when the mass condition $m_A > m_Z + m_S$ is satisfied. Given our DM mass benchmark, this threshold occurs at $m_A \approx 160$~GeV. Analogously, for the mono-$W$ channel, the on-shell production of charged scalars $h^\pm$ (followed by $h^\pm \to W^\pm S$) becomes kinematically allowed when $m_{h^\pm} > m_W + m_S \approx 155$~GeV.

\begin{figure}[htbp]
\centering
\includegraphics[width=0.4\textwidth]{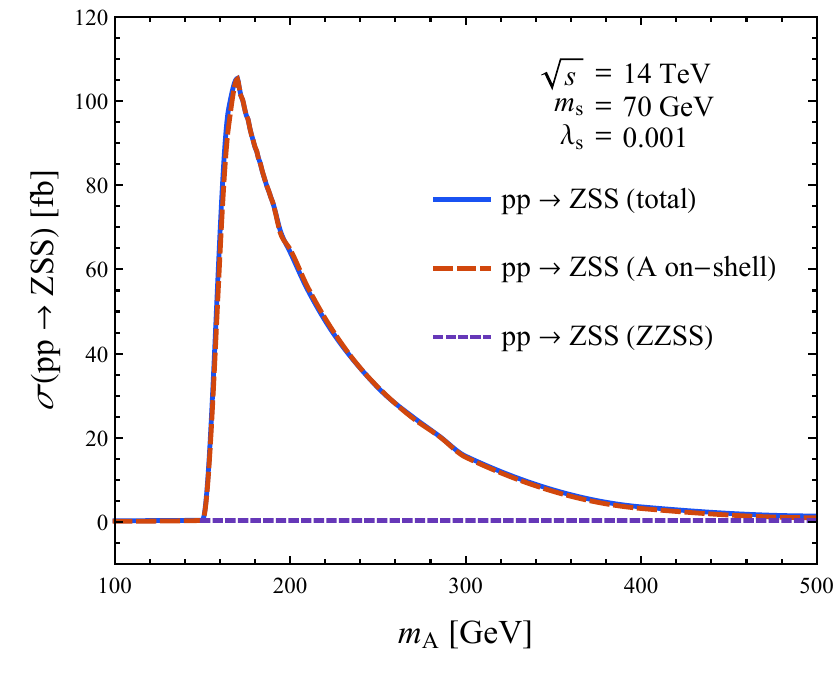}
\includegraphics[width=0.4\textwidth]{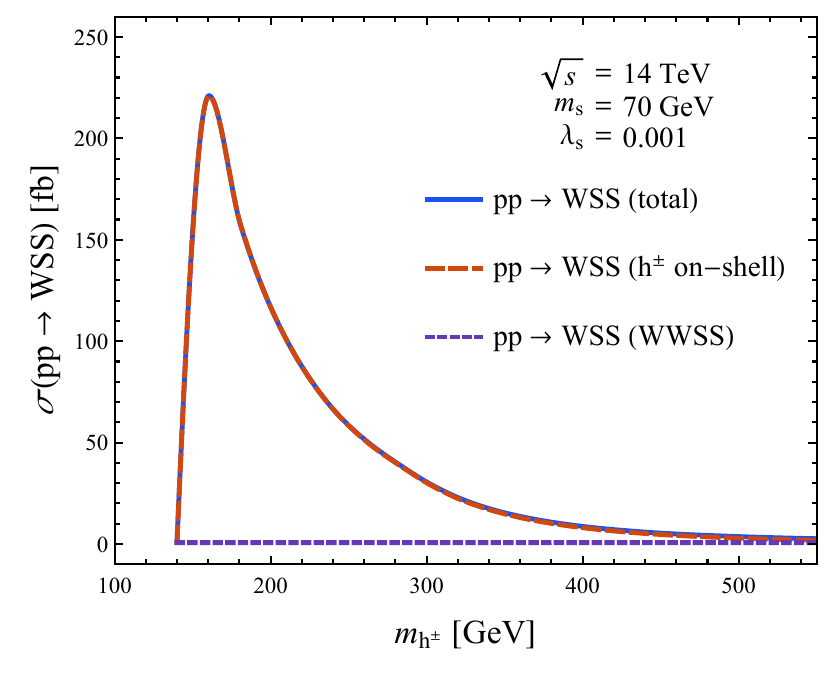}
\caption{Cross sections for mono-$Z$ and mono-$W$ associated DM production at $\sqrt{s}=14$ TeV as functions of $m_A$ (top) and $m_{h^\pm}$ (down). The plots demonstrate the transition from off-shell to on-shell dominance and the eventual suppression due to parton luminosities. In addition to displaying the variation in total production cross sections for different $m_A$ and $m_{h^{\pm}}$, we also separately show the contributions from on-shell $A$ or $h^{\pm}$ decays and from quartic $VVSS$ interactions for comparison. }
\label{fig:MA-pp_vss}
\end{figure}

This threshold effect is illustrated in Figure~\ref{fig:MA-pp_vss}. For both signal channels, the cross-section undergoes an enhancement as the mass of the parent particle ($A$ or $h^\pm$) crosses the on-shell threshold. These curves rapidly converge to the total cross-section above the threshold, confirming that the resonance-mediated topology becomes the dominant production mode in this regime. Conversely, the purple dashed lines denote the contribution of the $pp \to VSS$ process induced by the quartic $VVSS$ vertex. Compared to the on-shell $A$ or $h^\pm$ production, this contact-term contribution is negligible across the considered parameter space.

However, as the masses $m_A$ or $m_{h^\pm}$ continue to increase beyond the threshold, the signal cross-sections eventually begin to decrease. This decline is attributed to the suppression of parton luminosities at high $x$ values; as the required center-of-mass energy to produce heavier intermediate states increases, the probability of finding partons with sufficient momentum fractions in the colliding protons drops sharply. This competition between the phase-space enhancement at the kinematic turn-on and the subsequent PDF suppression defines the optimal search region for these inert scalars at the LHC.

\section{Signal-background Analysis and Significance}
\label{sec:result}

All Monte Carlo event generation for the signal and background processes is performed by \texttt{MadGraph5\_aMC@NLO}~\cite{madgraph}. Parton showering and hadronization are handled by \textsc{Pythia}\,8~\cite{pythia}, followed by a fast detector simulation utilizing \textsc{Delphes}\,3~\cite{delphes} with the default settings. The signal and SM background events are simulated in leading order. In the following subsections, we further separate our analysis into leptonic and hadronic channels. For the leptonic channel ($\ell^+\ell^- + E_{\text{T}}^{\text{miss}}$), we focus specifically on the mono-$Z$ case, as we have confirmed that distinguishing the mono-$W$ signals from the relevant backgrounds is very challenging. However, for the hadronic channel ($jj + E_{\text{T}}^{\text{miss}}$), both the mono-$Z$ and mono-$W$ cases are considered.

\subsection{The Leptonic Channel: \texorpdfstring{$\ell^+\ell^- + E_{\text{T}}^{\text{miss}}$}{ll+missing}}

The mono-$Z$ leptonic channel ($pp \to Z S S \to \ell^+\ell^- S S$) offers a remarkably clean experimental signature. The dominant irreducible backgrounds arise from diboson production, specifically $ZZ \to \ell^+\ell^-\nu\bar{\nu}$ and $W^+W^- \to \ell^+\nu\ell^-\bar{\nu}$. In addition, we carefully consider the reducible backgrounds in which misidentified particles generated $E_{\text{T}}^{\text{miss}}$. These include $WZ$ production, $Z+\text{jets}$, triboson production, $t\bar{t}$ production with leptonic decay, and Drell-Yan processes.

The dileptonic $t\bar{t}$ process constitutes an major irreducible background for the $\ell^+\ell^- + E_{\text{T}}^{\text{miss}}$ search. Despite the application of a $b$-jet veto to suppress top-quark contributions, a non-negligible fraction of $t\bar{t}$ events contaminate the signal region due to the finite $b$-tagging efficiency (typically $\sim 70\%$) and limited kinematic acceptance of the jets~\cite{ATLAS:2019bwq}. Furthermore, the presence of two neutrinos in the final state produces substantial real $E_{\text{T}}^{\text{miss}}$, mimicking the missing momentum signature of the DM candidates. 
The $W^\pm Z$ diboson production emerges as a significant background in the $\ell^+\ell^- + E_{\text{T}}^{\text{miss}}$ channel, primarily when one of the three final-state leptons fails to meet the identification or kinematic requirements.  We simulate such 'lost leptons' by assuming that the transverse momentum $p_{\text{T}}$ of the third-hardest lepton is sufficiently soft. Given typical LHC lepton acceptance rates, this misidentified background accounts for approximately $3\%$ of the total $W^\pm Z$ production~\cite{CMS:2017zts}. 
Other backgrounds from Drell-Yan, and triboson processes are found to be subdominant after accounting for LHC mis-tagging rates. Consequently, the major background processes for this channel focus on the irreducible $\ell\ell\nu\bar{\nu}$ and the dominant misidentified $t\bar{t}$ and $W^{\pm}Z$ background. 

We impose basic cuts requiring exactly two opposite-sign, same-flavor (OSSF) leptons with $p_{\text{T}}^\ell > 20$~GeV and pseudorapidity $|\eta_\ell| < 2.5$. Furthermore, since the $E_{\text{T}}^{\text{miss}}$ in parts of the background originates from unmeasured particles, its spectrum is inherently soft. Therefore, an initial requirement of $E_{\text{T}}^{\text{miss}} > 20$~GeV is applied at the baseline level to heavily suppress these mis-tagging SM backgrounds.

\begin{figure}[ht]
\centering
\includegraphics[width=0.45\textwidth]{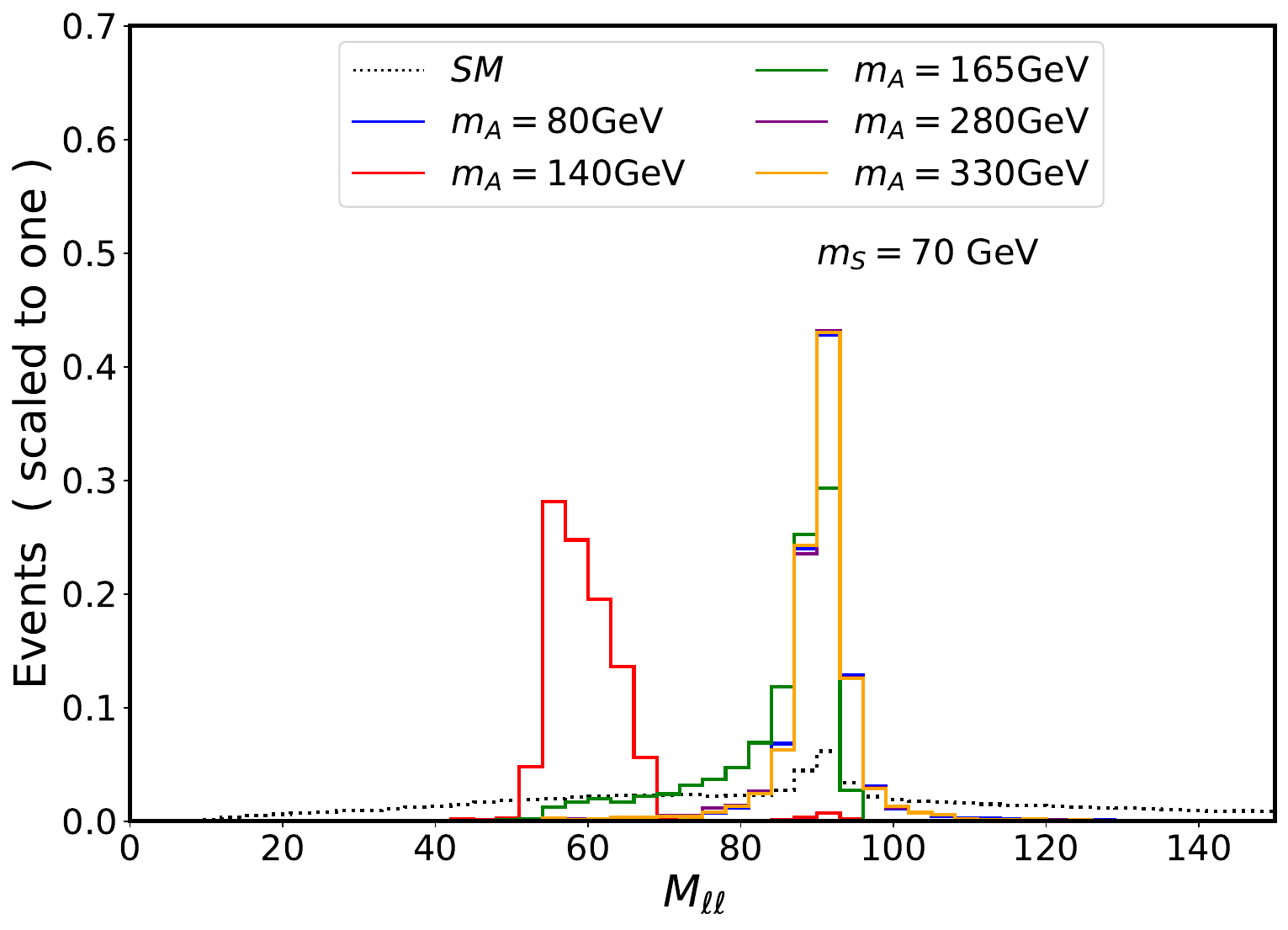}
\caption{Normalized invariant mass distributions of the dilepton system ($M_{\ell\ell}$) for the signals (solid colored lines) and the SM background (black dashed line). Signal results are shown for $m_A = 80$ (blue), 140 (red), 165 (green), 280 (purple), and 330 GeV (orange). }
\label{fig:mzll-met}
\end{figure}

To further suppress the combined SM background events, we exploited the distinctive kinematic features of the intermediate resonances. As discussed in the previous section, when the pseudoscalar mass exceeds the kinematic threshold ($m_A > m_Z+m_S$), the signal is dominated by the on-shell process $pp \to SA(\to ZS)$. Otherwise, it is mediated by an off-shell $Z$ boson. Consequently, the invariant mass of the lepton pair $M_{\ell\ell}$ serves as a powerful discriminant. Figure~\ref{fig:mzll-met} displays the $M_{\ell\ell}$ distributions taking $m_S = 70$~GeV as a benchmark. For $m_A > m_Z + m_S$, the signal is distinguished by a sharp peak at $M_Z$, whereas for $m_A < m_Z + m_S$, the off-shell effect broadens the distribution and shifts the peak toward lower mass values. In contrast, the combined SM background remains relatively smooth throughout the entire mass range. We thus employ two kinds of mass window based on the characteristic of the parameter space: $60~\text{GeV} < M_{\ell\ell} < 80~\text{GeV}$ for $m_A < 160$~GeV, and $80~\text{GeV} < M_{\ell\ell} < 100~\text{GeV}$ for $m_A > 160$~GeV.

\begin{figure}[htb]
\centering
\includegraphics[width=0.4\textwidth]{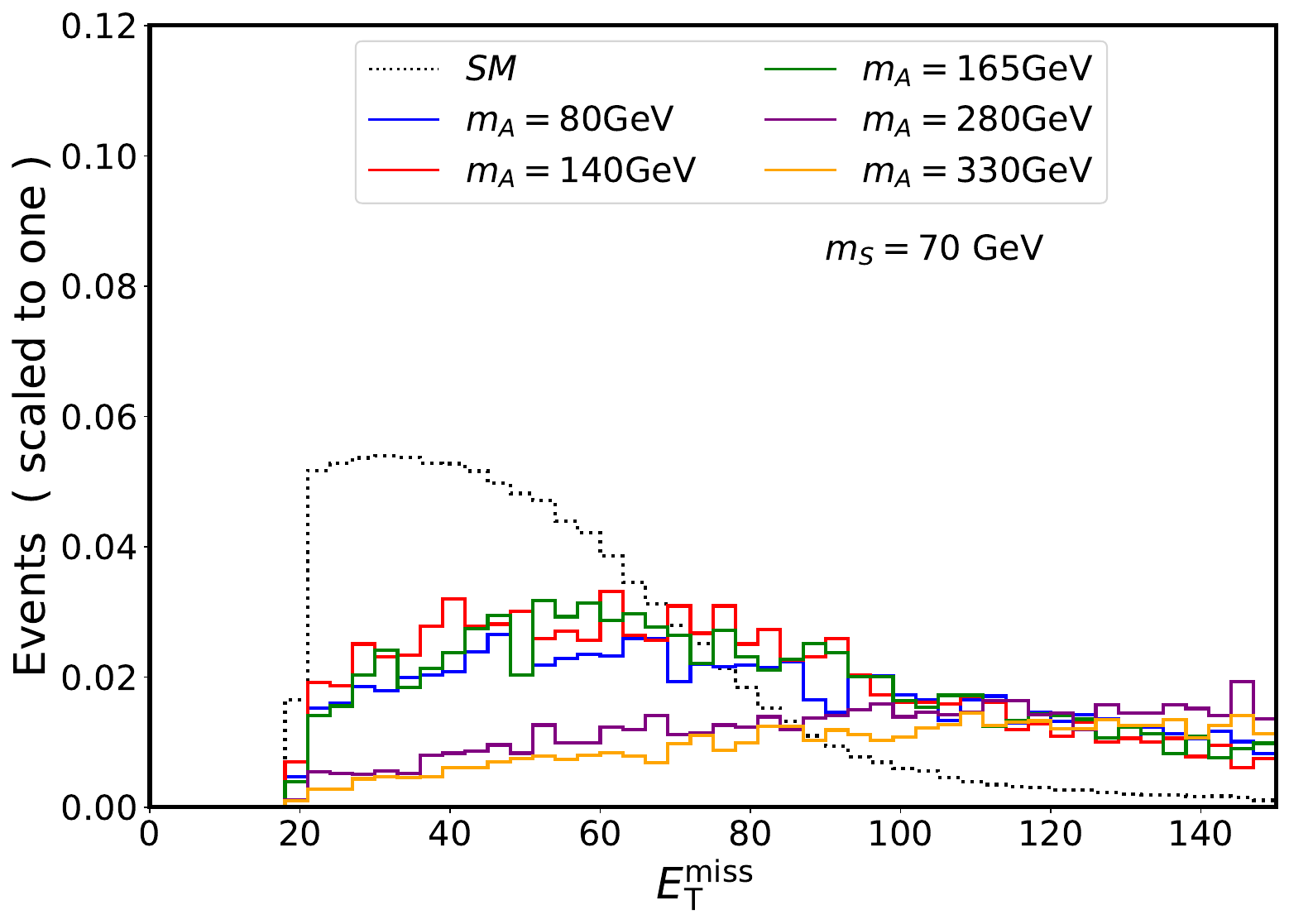} 
\includegraphics[width=0.4\textwidth]{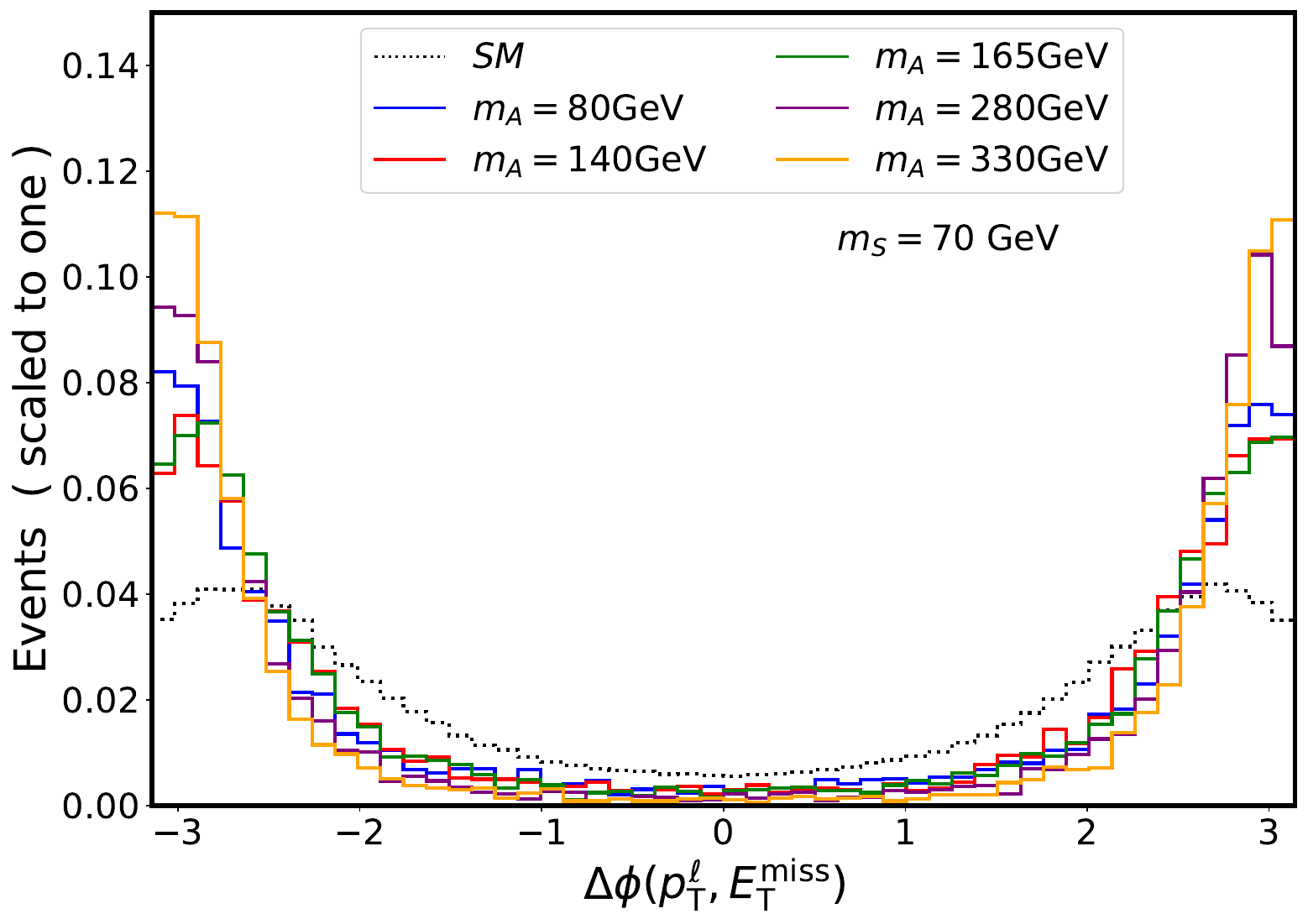}
\caption{Normalized distributions of the missing transverse momentum $E_{\text{T}}^{\text{miss}}$ (top) and the azimuthal angle difference $\Delta\phi(p_{\text{T}}^\ell, E_{\text{T}}^{\text{miss}})$ (down). The labels for the signal and background processes are the same as those in Figure~\ref{fig:mzll-met}.}
\label{fig:Deltaphi}
\end{figure}

Moreover, a mass splitting within the dark sector imparts a significant transverse boost to the invisible system. This results in a harder $E_{\text{T}}^{\text{miss}}$ spectrum for the signal and a highly back-to-back topology between the visible dilepton system and the missing momentum (see Figure~\ref{fig:Deltaphi}). Specifically, as $m_A$ increases beyond the threshold, the larger mass splitting $\Delta^0$ injects more kinetic energy into the $A \to ZS$ decay products, shifting the $E_{\text{T}}^{\text{miss}}$ distribution to higher energies. Below the threshold, phase-space suppression confines the $E_{\text{T}}^{\text{miss}}$ to soft regime, though it remains notably harder than the SM backgrounds. The azimuthal angle difference, $\Delta\phi(p_{\text{T}}^\ell, E_{\text{T}}^{\text{miss}})$, strongly peaks near $\pm\pi$, indicating that the lepton candidate and the DM pair recoil against each other. Accordingly, we introduce $E_{\text{T}}^{\text{miss}} > 60$~GeV and $|\Delta\phi| > 1.8$ as stringent optimization cuts.

\begin{figure}[ht]
\centering
\includegraphics[width=0.4\textwidth]{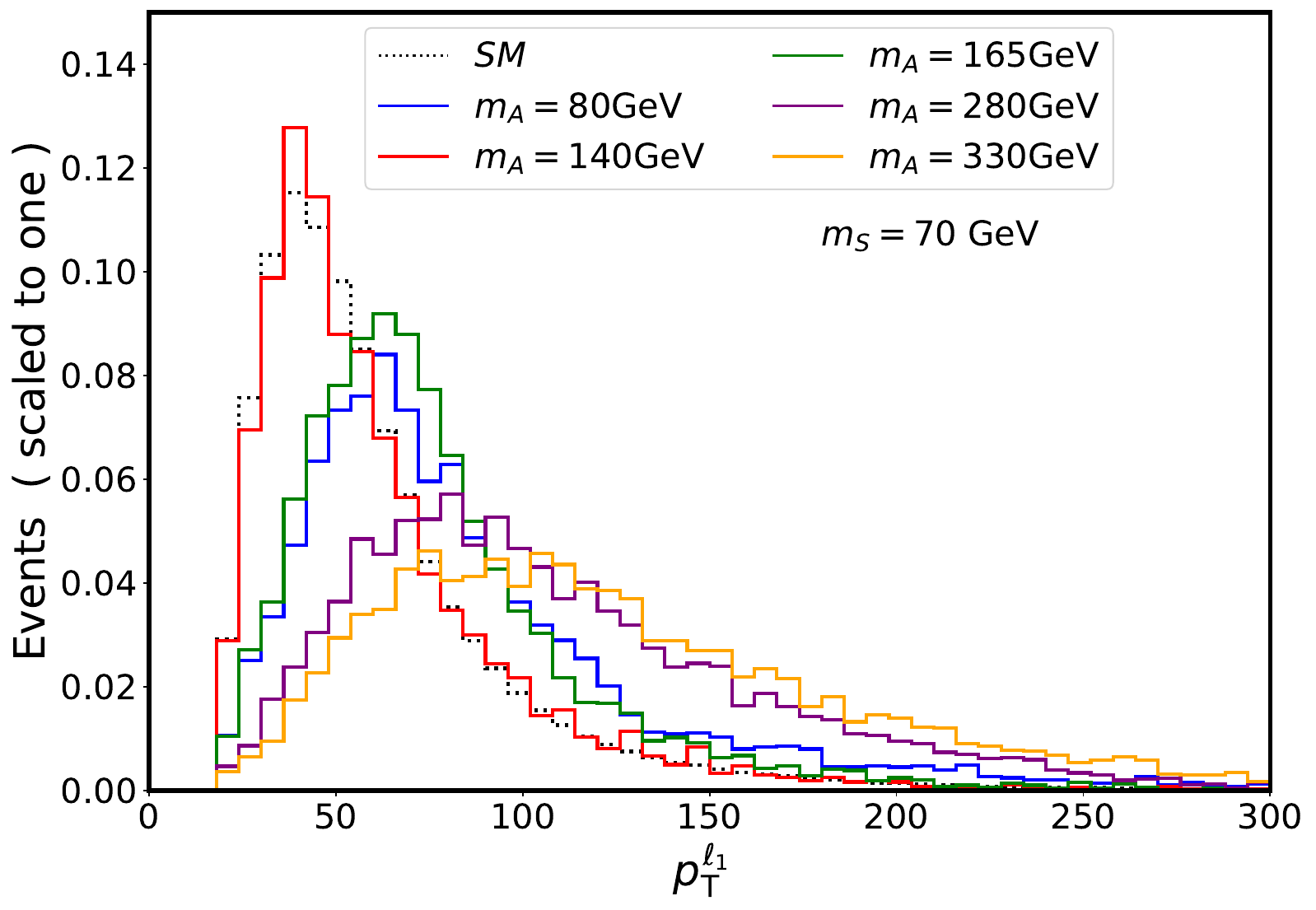}
\includegraphics[width=0.4\textwidth]{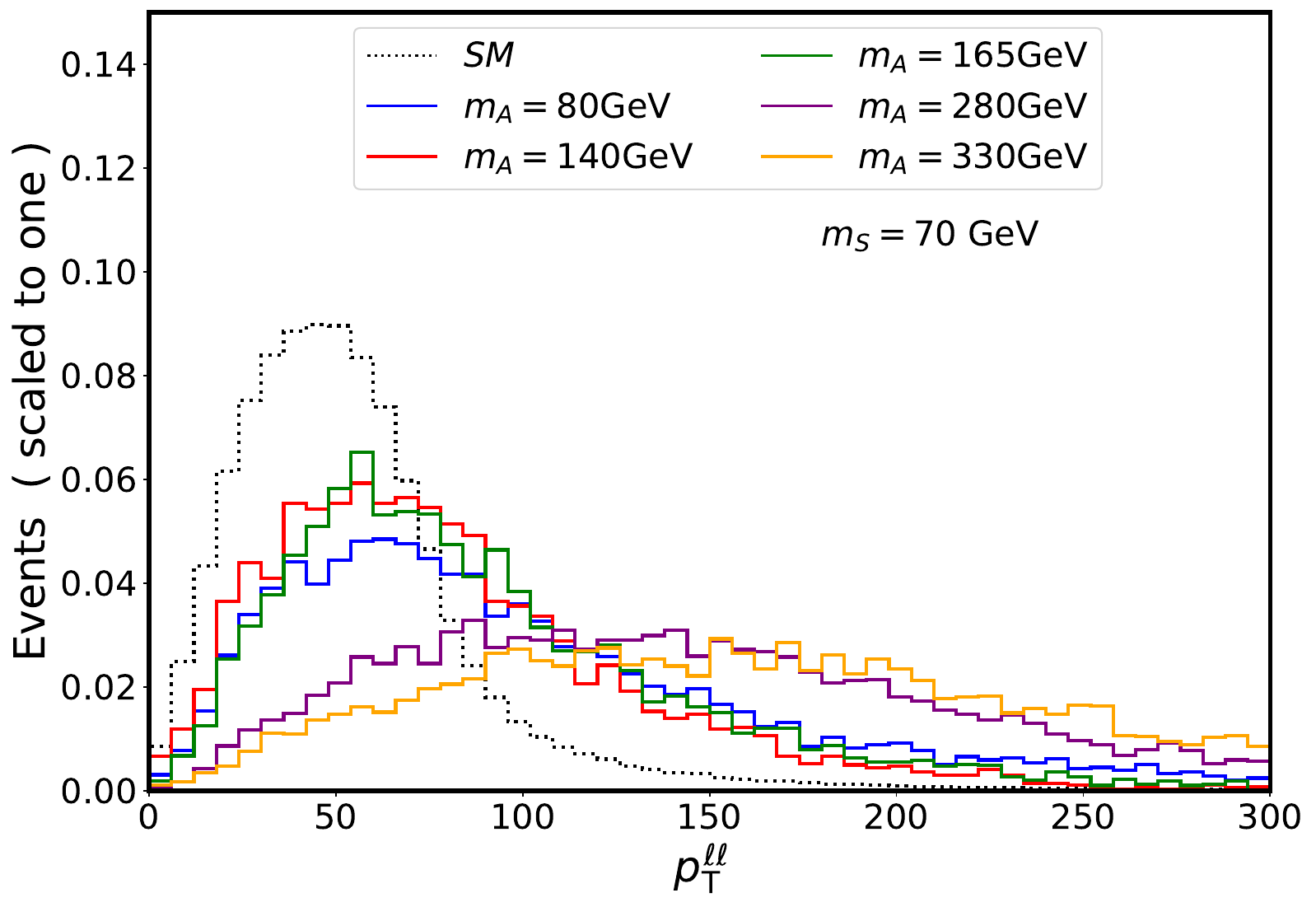}
\caption{Normalized distributions of the leading lepton transverse momentum $p_{\text{T}}^{\ell_1}$ (top) and the reconstructed $Z$ boson transverse momentum $p_{\text{T}}^{\ell\ell}$ (down). The labels for the signal and background processes are the same as those in Figure~\ref{fig:mzll-met}.}
\label{fig:ptl-ptzll}
\end{figure}

Figure~\ref{fig:ptl-ptzll} demonstrates that the transverse momenta of the leading lepton ($p_{\text{T}}^{\ell_1}$) and the reconstructed $Z$ boson ($p_{\text{T}}^{\ell\ell}$) exhibit similar behaviors, both increasing along with $m_A$. This is a direct consequence of the on-shell $A \to ZS$ decay kinematics, where a larger mass splitting provides a stronger recoil to the $Z$ boson. To leverage this signal advantage in the high-$p_{\text{T}}$ tail, we require $p_{\text{T}}^{\ell_1} > 50$~GeV and $p_{\text{T}}^{\ell\ell} > 60$~GeV.

The cut-flow evolution for the signal (assuming $m_S=70$~GeV and $\lambda_S = 10^{-3}$) and three major backgrounds at $\sqrt{s}=14$~TeV are summarized in Table~\ref{tab:allCut}. Comparing the survival rates across different benchmarks reveals that the LHC achieves optimal sensitivity to the DM signal near the on-shell resonance threshold ($m_A \approx 160$~GeV).

\begin{table*}[htbp]
\centering
\caption{Cut-flow cross-sections (in fb) for the leptonic channel at $\sqrt{s} = 14$ TeV. The signal process $pp \to Z S S \to \ell^+\ell^- S S$ with $m_S=70$ GeV includes three benchmark $m_A$ values: 165, 200, 205 GeV. The major background processes are involved: Irreducible $\ell\ell\nu\bar{\nu}$, $t\bar{t}$, and $WZ$. }
\label{tab:allCut}
\vspace{4pt}
\begin{tabular}{lcccccc}
\toprule
\multirow{2}{*}{Cuts} & \multicolumn{3}{c}{Backgrounds} & \multicolumn{3}{c}{Signals ($m_S=70\ \rm{GeV},\, \lambda_S=10^{-3}$)} \\
\cmidrule(lr){2-4} \cmidrule(lr){5-7}
                   & $\ell\ell\nu\bar{\nu}$ & $t\bar{t}$ & $WZ$  & $m_A=165\ \rm{GeV}$ & $m_A=200\ \rm{GeV}$ & $m_A=205\ \rm{GeV}$ \\
\midrule
Basic cut                              & 794.36 & 338.86 & 0.91 &   3.90   &   1.65   & 1.55 \\
$E_{\text{T}}^{\text{miss}} > 60$ GeV  & 261.99 & 200.50 & 0.32  &   2.69   &   1.30   & 1.21 \\
$p_{\text{T}}^{\ell_1} > 50$ GeV       & 163.41 & 133.41 & 0.24   &   2.28   &   1.17   & 1.11 \\
$80 < M_{\ell\ell} < 100$ GeV          & 46.99  & 16.02 & 0.14   &   2.28   &   1.17   & 1.11 \\
$p_{\text{T}}^{\ell\ell} > 60$ GeV     & 36.44  & 10.46 & 0.10    &   1.60   &   1.00   & 0.95 \\
$|\Delta\phi| > 1.8$                   & 32.29  & 6.21 & 0.07   &   1.50   &   0.95   & 0.91 \\
\bottomrule
\end{tabular}
\end{table*}

\begin{figure}[htbp]
\centering
\includegraphics[width=0.4\textwidth]{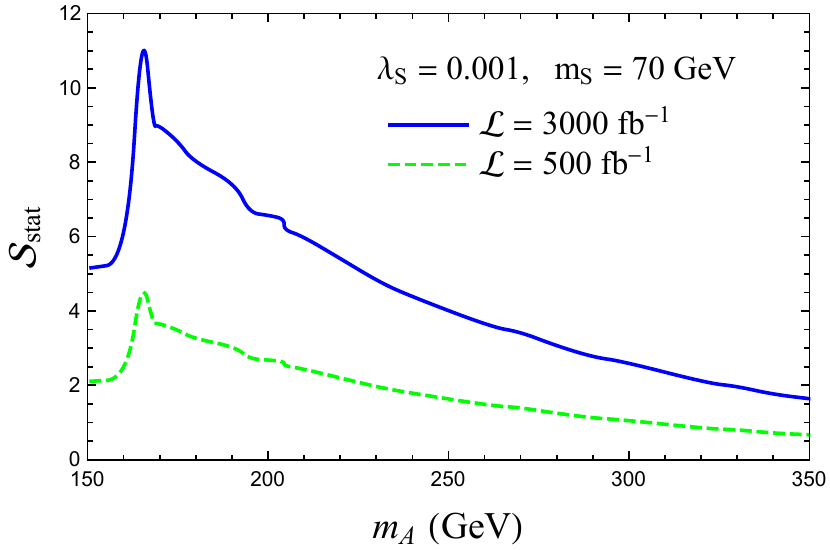}
\caption{Expected statistical significance $\mathcal{S}_{\text{stat}}$ for the leptonic channel ($pp \to Z S S \to \ell^+\ell^- S S$) at $\sqrt{s} = 14$ TeV as a function of the pseudoscalar mass $m_A$, assuming integrated luminosities of $500~\text{fb}^{-1}$ and $3000~\text{fb}^{-1}$.}
\label{fig:llvv-ss=2}
\end{figure}

Based on this event selection strategy, we calculate the expected statistical significance~\cite{Cowan:2010js},
\begin{equation}
\mathcal{S}_{\text{stat}}
=
\sqrt{\,2\left[(N_{\text{bg}}+N_s)\ln\!\left(1+\frac{N_s}{N_{\text{bg}}}\right)-N_s\right]}\, ,
\end{equation}
where $N_s$ and $N_{\text{bg}}$ are the numbers of signal and background events expected after all cuts. 
As shown in Figure~\ref{fig:llvv-ss=2}, the significance peaks sharply near $m_A \approx 165$~GeV, corresponding to the kinematic unblocking of the $A \to ZS$ decay channel. At an integrated luminosity of $\mathcal{L} = 500~\text{fb}^{-1}$, the $2\sigma$ exclusion limit is approximately $m_A \sim 230$~GeV. With the High-Luminosity LHC (HL-LHC) reaching $\mathcal{L} = 3000~\text{fb}^{-1}$, the discovery potential for this resonance-enhanced parameter space can be extended to approximately $m_A \sim 330$~GeV.

\subsection{The Hadronic Channel: \texorpdfstring{$jj + E_{\text{T}}^{\text{miss}}$}{jj+missing}}

The hadronic channel ($pp \to jj S S$) consolidates contributions from both $Z \to jj$ and $W^\pm \to jj$ decays of the signal processes shown in Figure~\ref{fig:feynman}. The primary irreducible background is the $Z(\to \nu\bar{\nu}) + \text{jets}$ process. Furthermore, semi-leptonic $W + \text{jets}$ ($W \to \ell\nu$) events constitutes a prominent background, as kinematic acceptance losses or lepton identification inefficiencies can cause them to identically fake the $jj + E_{\text{T}}^{\text{miss}}$ final state. To prevent double-counting between the matrix element and the parton shower, the MLM matching scheme~\cite{Mangano:2006rw} is applied with the merging scale set to ${\rm xqcut} = 50~\text{GeV}$ for these two multijet processes. Thus, the major background processes for the hadronic channel focuses on the irreducible QCD multijet with missing energy $jj\nu\bar{\nu}$, and the misidentified $W + \text{jets}$ backgrounds.

\begin{figure}[htbp]
\centering
\includegraphics[width=0.4\textwidth]{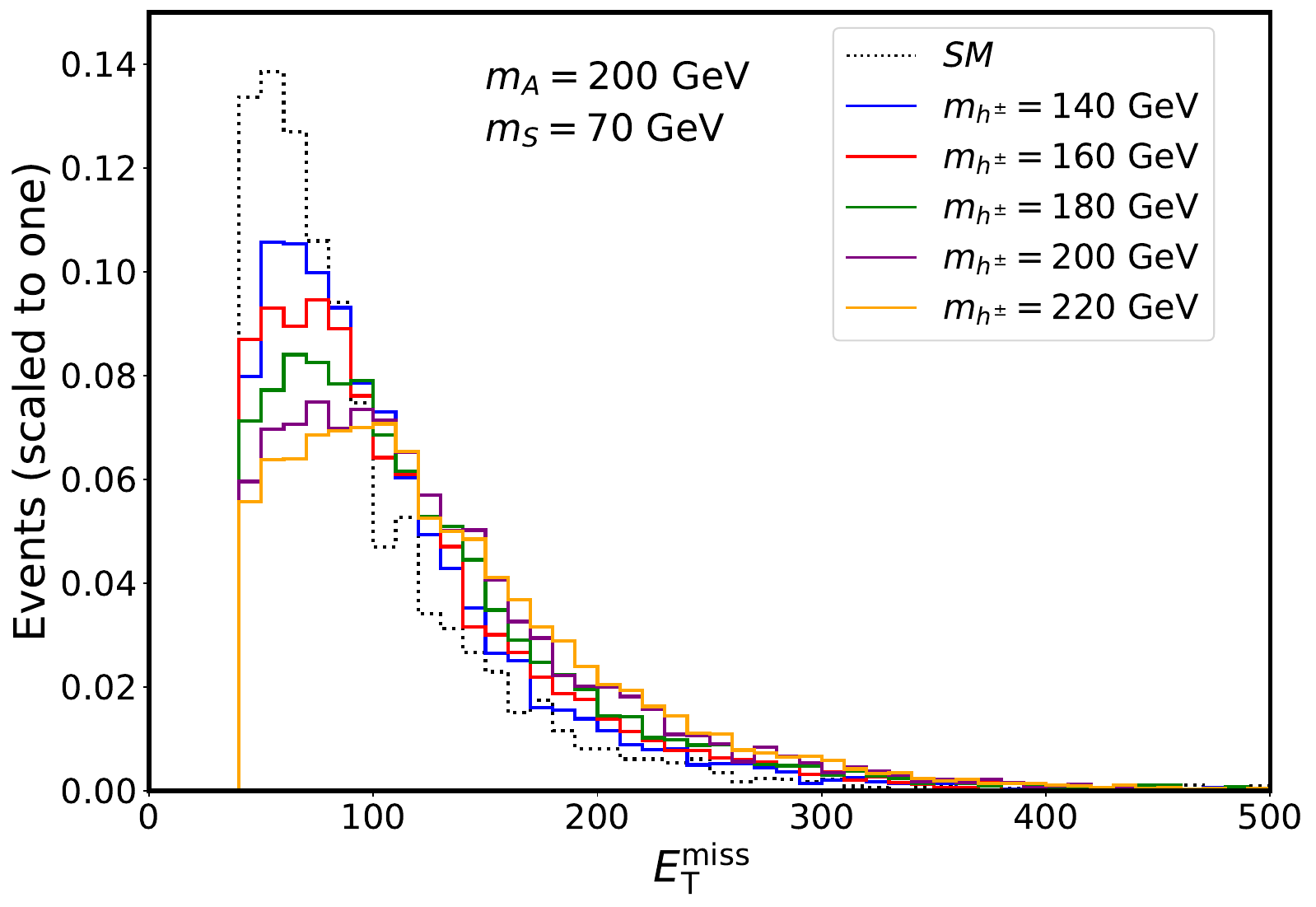}
\includegraphics[width=0.4\textwidth]{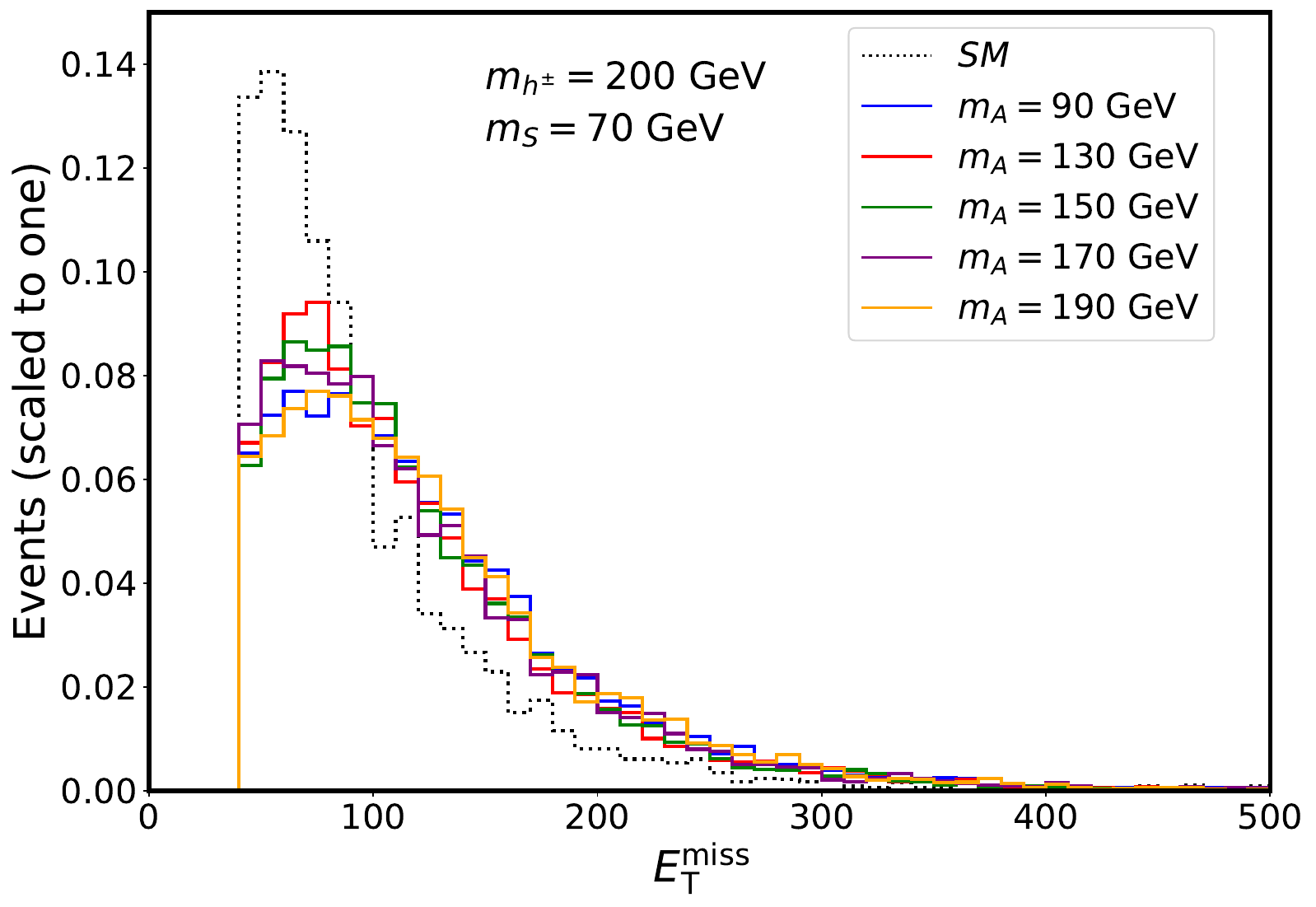}
\caption{Normalized distributions of the missing transverse momentum $E_{\text{T}}^{\text{miss}}$ for the $pp \to W^{\pm} S S \to jj S S$ (top) and $pp \to Z S S \to jj S S$ (down). The benchmark values of $m_{h^{\pm}}$ ($m_A$) are labeled as 140 (90) GeV with blue, 160 (130) GeV with red, 180 (150) GeV with green, 200 (170) GeV with purple, and 220 (190) GeV with orange. }
\label{fig:MjjMET}
\end{figure}

We establish a benchmark at $\sqrt{s}=14$~TeV with $m_S=70$~GeV, $m_{A,{h^\pm}} = 200$~GeV, and $\lambda_S=10^{-3}$ to analyze the kinematic distributions of the hadronic channel. 
To efficiently suppress overwhelming QCD multi-jet background, we implement an adaptive kinematic selection strategy that dynamically responds to the resonance features. The basic cuts require: 
\[ N_{\text{jet}} \geq 2, \, p_{\text{T}}^{\text{jet}} > 20~\text{GeV}, \, |\eta_{\text{jet}}| < 2.5, \, E_{\text{T}}^{\text{miss}} > 40~\text{GeV}.\] 
Since the signal contains two stable inert scalars $S$ that carry away substantial transverse momentum, its $E_{\text{T}}^{\text{miss}}$ distribution is significantly harder than that of the SM backgrounds (Figure~\ref{fig:MjjMET}). We therefore impose an aggressive cut of $E_{\text{T}}^{\text{miss}} > 100$~GeV. Note that while extreme $E_{\text{T}}^{\text{miss}}$ can lead to highly boosted $W/Z$ bosons whose decay products merge into fat-jets, the recoil $p_{\text{T}}$ in our model predominantly clusters around 100~GeV. Consequently, resolved di-jet analysis is sufficient, and jet substructure techniques are not required in this study.

\begin{figure}
\centering
\includegraphics[width=0.4\textwidth]{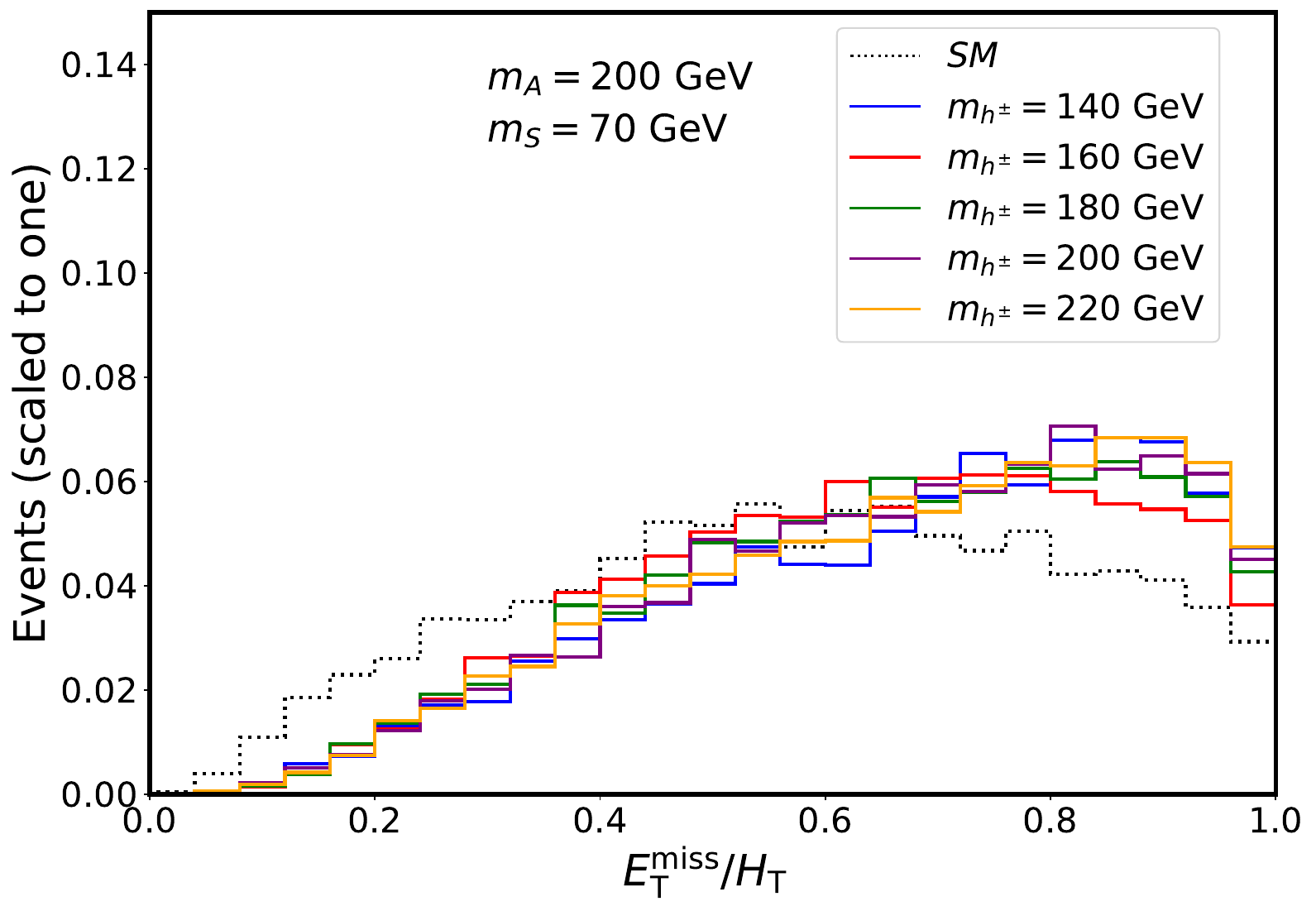}
\includegraphics[width=0.4\textwidth]{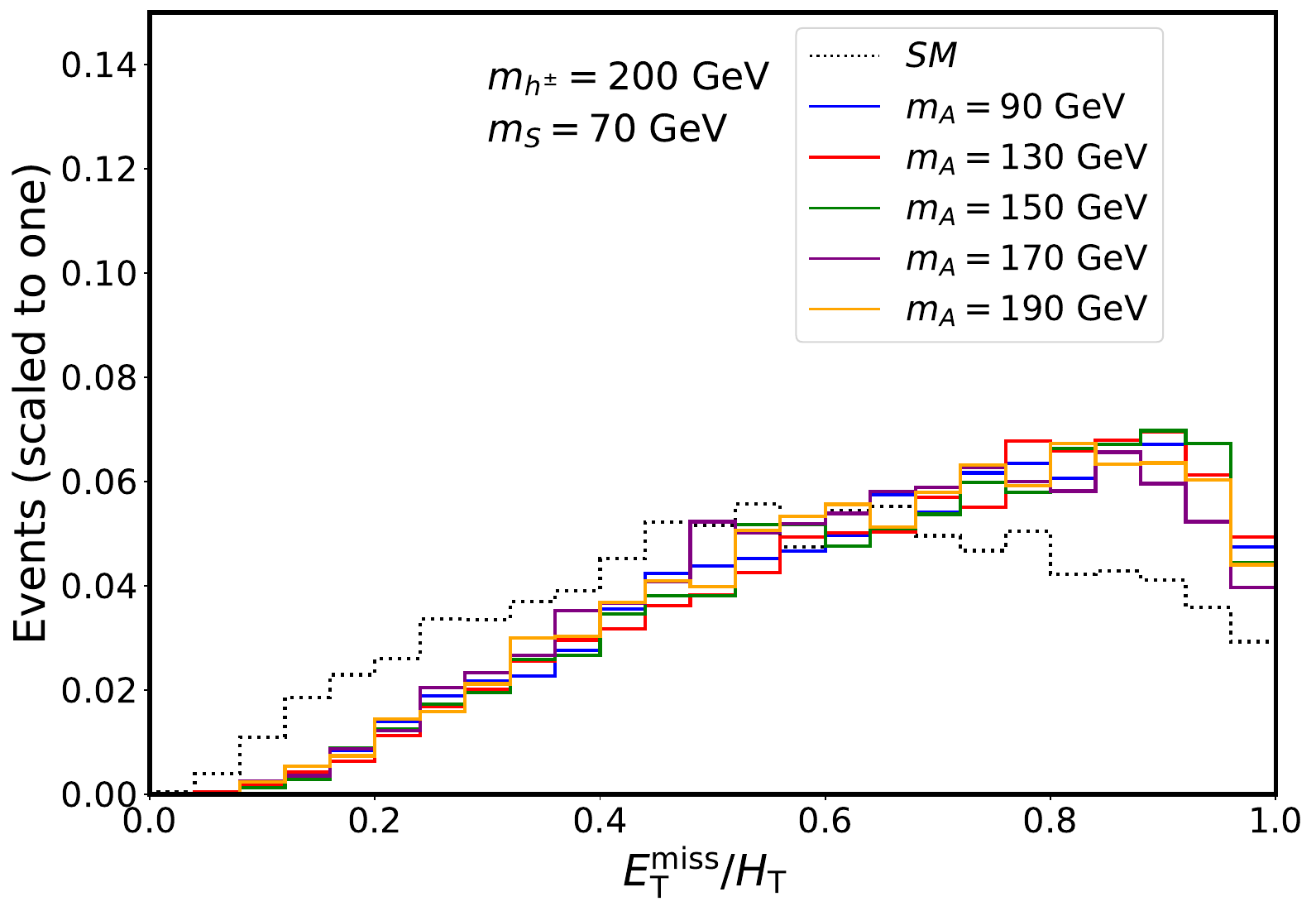}
\caption{Normalized distributions of the transverse energy ratio $E_{\text{T}}^{\text{miss}}/H_{\text{T}}$ for the $pp \to W^{\pm} S S \to jj S S$ (top) and $pp \to Z S S \to jj S S$ (down). The labels for the signal and background processes are the same as those in Figure~\ref{fig:MjjMET}.}  
\label{fig:METHT}
\end{figure}

Furthermore, we consider the missing transverse momentum, $E_{\text{T}}^{\text{miss}}$, and the scalar sum of the transverse momenta of all final-state jets, $H_{\text{T}} = \sum_{i} |p_{\text{T},i}|$. To effectively discriminate the IDM signal from SM backgrounds, we introduce the dimensionless ratio $E_{\text{T}}^{\text{miss}} / H_{\text{T}}$. By construction, $E_{\text{T}}^{\text{miss}} / H_{\text{T}} \le 1$ since the magnitude of the vector sum of transverse momenta is bounded by their scalar sum.  
For the signal process, the DM particles originate from the hard scattering and carry a substantial fraction of the event's transverse momentum, resulting in a higher $E_{\text{T}}^{\text{miss}} / H_{\text{T}}$ value. 
In contrast, the jets in $W/Z + \text{jets}$ processes are predominantly produced by initial state radiation (ISR), where $H_{\text{T}}$ scales more rapidly than the genuine $E_{\text{T}}^{\text{miss}}$. As illustrated in Figure~\ref{fig:METHT}, a requirement of $E_{\text{T}}^{\text{miss}} / H_{\text{T}} > 0.6$ can significantly suppress the SM backgrounds.

\begin{figure}[htbp]
\vspace{2pt}
\centering
\includegraphics[width=0.4\textwidth]{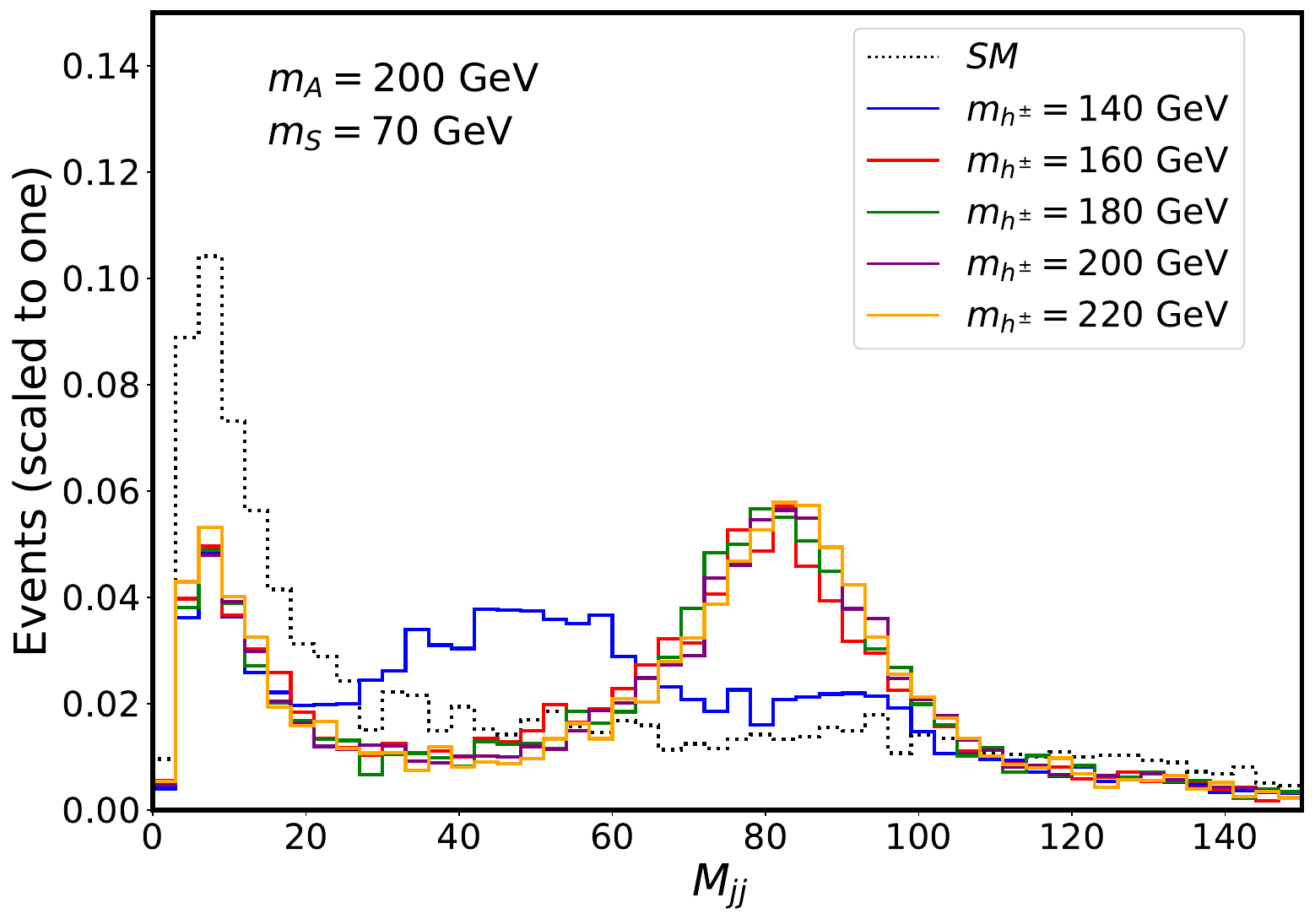}
\includegraphics[width=0.4\textwidth]{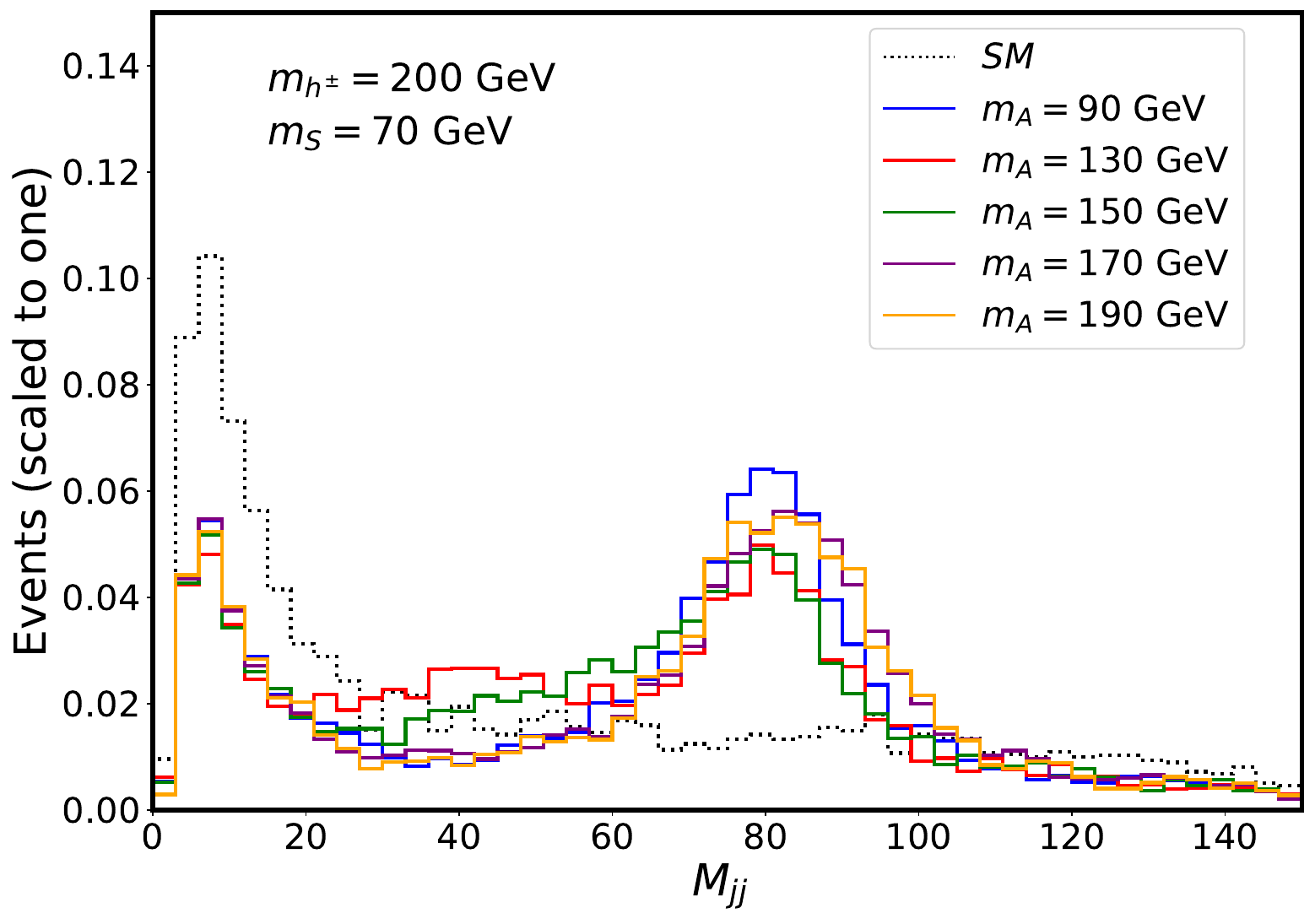} 
\caption{Normalized distributions of the di-jet invariant mass $M_{jj}$ for the $pp \to W^{\pm} S S \to jj S S$ (top) and $pp \to Z S S \to jj S S$ (down). The labels for the signal and background processes are the same as those in Figure~\ref{fig:MjjMET}.} 
\label{fig:jjss-mjj}
\end{figure}

Depending on whether the intermediate scalar masses $m_A$ and $m_{h^\pm}$ exceed their respective kinematic thresholds, the morphology of the $M_{jj}$ distribution undergoes significant changes, as shown in Figure~\ref{fig:jjss-mjj}. 
The parameter space is primarily bifurcated by the $h^\pm \to W^\pm S$ decay threshold. Above $m_{h^\pm} \approx 150$~GeV, the on-shell production of $W$ bosons yields a distinct mass pole in the $M_{jj}$ distribution. Below this threshold, the process is governed by off-shell $W^\ast$ exchange, resulting in a broad and flattened continuum that necessitates an alternative selection window to effectively capture the signal. 
Furthermore, the $M_{jj}$ profile is modulated by $m_A$, which we categorize into three distinct kinematic regimes. For $m_A < 120$~GeV, the $Z$ boson is deeply off-shell, and the signal relies almost entirely on the $W$ resonance from the charged sector. In the transition region ($120~\text{GeV} < m_A < 160~\text{GeV}$), the $Z^\ast$ contribution begins to harden, leading to a broader $M_{jj}$ profile. Once $m_A$ exceeds $160$~GeV, the $A \to ZS$ channel is kinematically unblocked, yielding a robust $Z$-boson resonance peak. 
It is important to note that even in the fully on-shell regimes, the $W$ and $Z$ mass poles are located in close proximity. Due to the finite jet energy resolution of the detector, these two peaks are often smeared into a single broad structure in the $M_{jj}$ spectrum. 

Consequently, our $M_{jj}$ selection window of $60\text{--}100$~GeV in the resonance-dominated regions is specifically designed to encompass both the $W$ and $Z$ contributions, thus ensuring optimal signal efficiency. To maximize the search sensitivity across the diverse parameter space, we employ the following adaptive $M_{jj}$ mass windows: 
\begin{itemize}
\item $20~\text{GeV} < M_{jj} < 80~\text{GeV}$,\quad for $m_{h^\pm} < 150~\text{GeV}$;
\item $60~\text{GeV} < M_{jj} < 100~\text{GeV}$,\quad for $m_{h^\pm} > 150~\text{GeV}$, 
      with $m_A < 120~\text{GeV}$ or $m_A > 160~\text{GeV}$;
\item $20~\text{GeV} < M_{jj} < 100~\text{GeV}$,\quad for $m_{h^\pm} > 150~\text{GeV}$, 
      with $120~\text{GeV} < m_A < 160~\text{GeV}$.
\end{itemize}
When $M_{jj}$ falls within the $W/Z$ mass window, the kinematic resemblance between the signal and the irreducible background is maximized, thereby mandating the stringent synergy of the aforementioned $E_{\text{T}}^{\text{miss}}$ and $E_{\text{T}}^{\text{miss}}/H_{\text{T}}$ cuts.

\begin{table*}[htbp]
\caption{Cut-flow cross-sections (in fb) for the hadronic channel at $\sqrt{s} = 14$ TeV. Three signal benchmark points $(m_A, m_{h^{\pm}})$ are chosen for illustration. The combined backgrounds consist of $W/Z + \text{jets}$.}
\label{tab:jjss-Cutflow}
\centering
\begin{tabular}{lcccc}
\toprule
\multirow{3}{*}{Cuts} & \multirow{3}{*}{Backgrounds} & \multicolumn{3}{c}{Signals ($m_S=70\ \rm{GeV},\, \lambda_S=10^{-3}$)} \\
\cmidrule(lr){3-5}
 &  & \makecell{$m_{A}=200~\mathrm{GeV}$\\$m_{h^\pm}=200~\mathrm{GeV}$} & \makecell{$m_{A}=200~\mathrm{GeV}$\\$m_{h^\pm}=180~\mathrm{GeV}$} & \makecell{$m_{A}=170~\mathrm{GeV}$\\$m_{h^\pm}=200~\mathrm{GeV}$} \\
\midrule
Basic Cut & 847773.62 & 62.23 & 73.20 & 74.89 \\
$E_{\text{T}}^{\text{miss}}>100~\mathrm{GeV}$ & 276355.31 & 36.22 & 38.61 & 39.09 \\
$E_{\text{T}}^{\text{miss}}/H_{\text{T}} > 0.6$ & 160960.30 & 28.92 & 30.55 & 30.74 \\
$60 < M_{jj} < 100~\mathrm{GeV}$ & 28524.61 & 16.59 & 17.66 & 17.53  \\
\bottomrule
\end{tabular}
\end{table*}

To illustrate the efficacy of this event selection strategy, Table~\ref{tab:jjss-Cutflow} details the cut-flow cross-sections for three representative signal benchmarks and the combination of $W/Z + \text{jets}$ background processes. We specifically highlight benchmark points situated in the fully on-shell resonance regime ($m_A \ge 170~\mathrm{GeV}$ and $m_{h^\pm} \ge 180~\mathrm{GeV}$), where the discovery potential is most pronounced. In strict accordance with our adaptive $M_{jj}$ mass windows, the $60 < M_{jj} < 100~\mathrm{GeV}$ is explicitly applied for these specific high-mass benchmarks.  Furthermore, we omit $p_{\text{T}}^{\text{jet}}$ and $\Delta\phi(p_{\text{T}}^{\text{jet}}, E_{\text{T}}^{\text{miss}})$ from the selection criteria in this channel, as their topological distributions provide negligible discriminating power between the signal and the SM background.

\begin{figure}[htbp]
\centering
\includegraphics[width=0.4\textwidth]{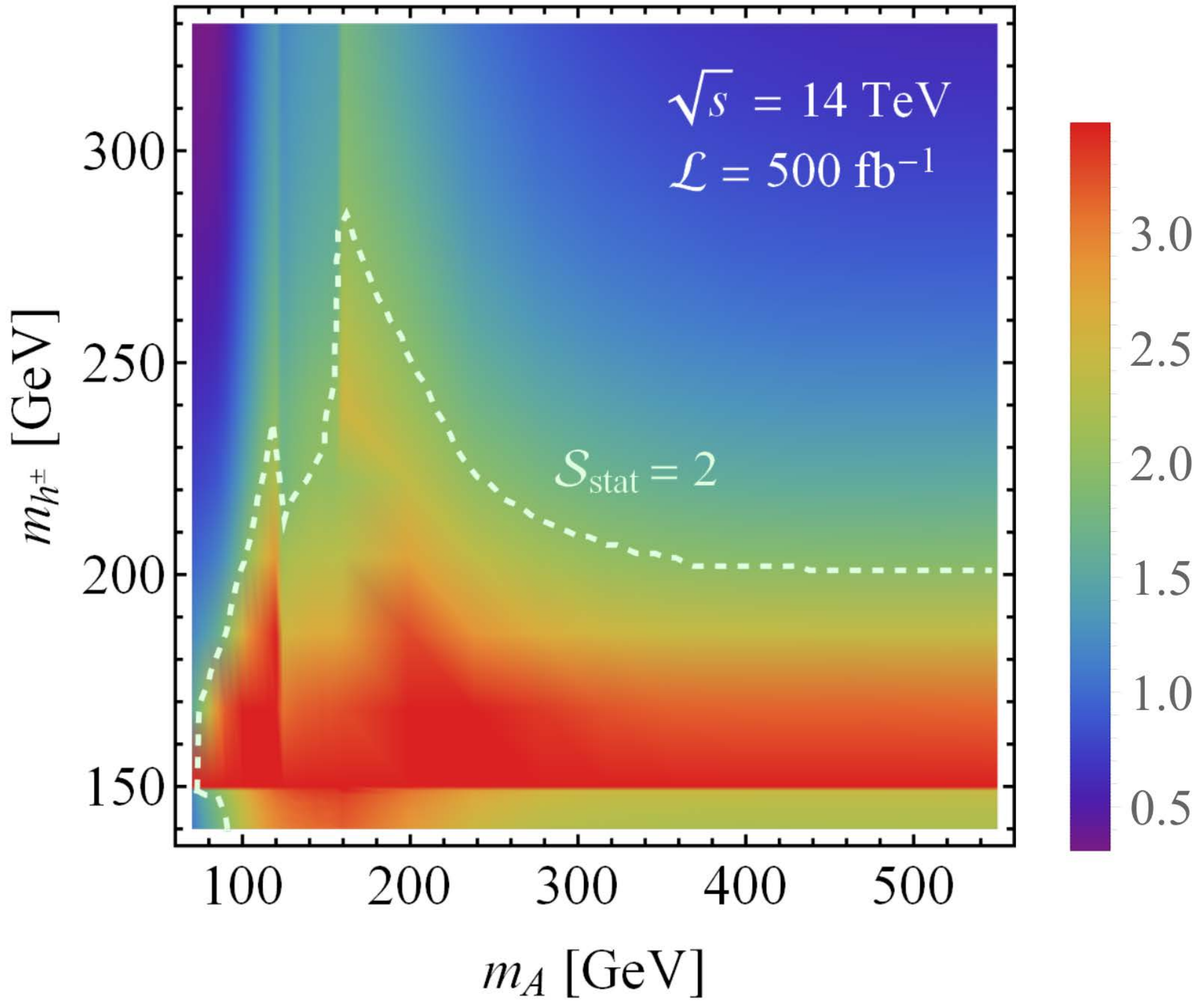}
\includegraphics[width=0.4\textwidth]{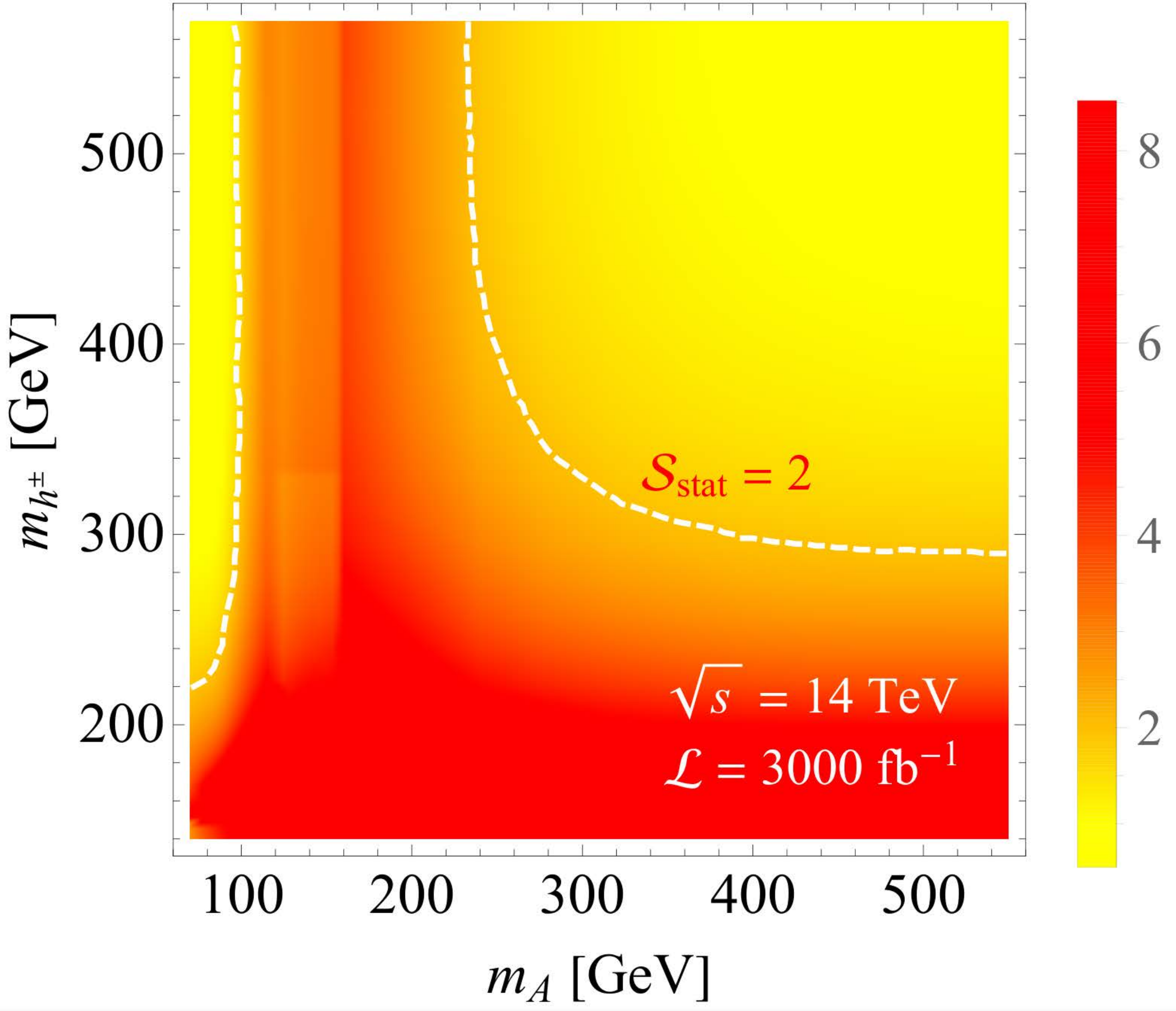}
\caption{Contours of statistical significance $\mathcal{S}_{\text{stat}}$ in the $(m_A, m_{h^\pm})$ plane for the hadronic channel ($pp \to jj S S$) at $\sqrt{s} = 14$ TeV with integrated luminosities of $\mathcal{L} = 500~\text{fb}^{-1}$ (top) and $\mathcal{L} = 3000~\text{fb}^{-1}$ (bottom).}
\label{fig:ss-jjl}
\end{figure}

The projected statistical significance $\mathcal{S}_{\text{stat}}$ for the hadronic channel ($jjSS$) in the $(m_A, m_{h^\pm})$ parameter plane is presented in Figure~\ref{fig:ss-jjl}, assuming a fixed DM mass $m_S = 70$~GeV. The results are shown for both the current LHC stage with $\mathcal{L} = 500~\text{fb}^{-1}$ and the future HL-LHC with $\mathcal{L} = 3000~\text{fb}^{-1}$. The dashed white contour delineates the $2\sigma$ exclusion boundary. A striking feature of the significance contours is the presence of discrete, step-like boundaries—most notably the sharp vertical discontinuities at $m_A \approx 110$ and $160$~GeV. These non-continuous transitions are not intrinsic to the underlying physics but are a direct consequence of the adaptive $M_{jj}$ mass window strategy.

At $\mathcal{L} = 500\ \text{fb}^{-1}$, the $jjSS$ channel can achieve a $2\sigma$ exclusion sensitivity up to $m_{h^{\pm}} \approx 280$ GeV, with the peak sensitivity occurring at $m_A \approx 190$ GeV. Moreover, for $m_{h^{\pm}} \lesssim 200$ GeV, the $2\sigma$ exclusion sensitivity can reach $m_A \gtrsim 550$ GeV. With the HL‑LHC upgrade to $\mathcal{L} = 3000\ \text{fb}^{-1}$, regions previously limited to $2\sigma$ significance are elevated to the $5\sigma$ discovery level. In particular, within the region defined by $100 \lesssim m_A \lesssim 220$ GeV, a significance exceeding $2\sigma$ can be achieved for $m_{h^{\pm}} \lesssim 550$ GeV. Similarly, for $m_{h^{\pm}} \lesssim 300$ GeV, the $2\sigma$ exclusion sensitivity can reach $100 \lesssim m_A \lesssim 550$ GeV. Therefore, a large portion of the IDM parameter space capable of explaining both the Fermi‑LAT GCE and the AMS‑02 antiproton excess (see Table~\ref{tab:para}) is likely to be detectable at the HL‑LHC.

\section{Conclusion}
\label{sec:conclusion}

In this work, we have comprehensively investigated the viability of the IDM as a unified explanation for the Fermi-LAT Galactic Center gamma-ray and AMS-02 antiproton excesses. Motivated by phenomenological global fits that tightly constrain the dark matter (DM) mass to the $54\text{--}74$~GeV window and strongly favor the gauge-dominated $SS \to WW^\ast$ annihilation mechanism to explain these two astrophyiscal excesses for the DM mass within $70\text{--}74$~GeV, we systematically evaluated the collider discovery prospects at the $\sqrt{s}=14$~TeV HL-LHC.

Our analysis demonstrates that the mono-$Z$ ($pp \to \ell^+\ell^- SS$) and mono-$W$ ($pp \to jj SS$) signatures exhibit a highly complementary sensitivity to the internal mass spectrum of the dark sector. The mono-$Z$ process serves as a targeted probe for the neutral mass splitting $\Delta^0$, whereas the mono-$W$ process independently maps out the charged scalar physics governing $\Delta^\pm$. Furthermore, the onset of on-shell resonance decay channels at the specific kinematic boundaries of $m_A > m_Z + m_S$ and $m_{h^\pm} > m_W + m_S$ significantly amplifies the production cross-sections to act as a powerful kinematic booster. 

Using our event selection strategy, we demonstrate that the HL-LHC operating at an integrated luminosity of $3000~\text{fb}^{-1}$ can achieve decisive statistical significance. The discovery potential peaks sharply near the resonance boundaries (e.g., $m_A \approx 165$~GeV), providing a clear and accessible experimental target. 
Our projections at an integrated luminosity of $\mathcal{L} = 500~\mathrm{fb}^{-1}$ indicate that the leptonic channel can achieve a $2\sigma$ exclusion sensitivity for pseudoscalar masses up to $m_A \approx 230$~GeV. As the luminosity is upgraded to $3~\mathrm{ab}^{-1}$ during the HL-LHC era, the $2\sigma$ exclusion sensitivity for this channel can expand to $m_A \approx 330$~GeV such that the detection of $80 \lesssim \Delta^0 \lesssim 260$ GeV will be covered. 

In comparison, the hadronic channel substantially broadens the overall coverage. At $\mathcal{L} = 500\ \mathrm{fb}^{-1}$, it extends the $2\sigma$ reach to probe charged scalar masses up to $m_{h^{\pm}} \approx 280$ GeV, with the best pseudoscalar detection sensitivity occurring at $m_A \approx 190$ GeV, thus covering a large fraction of the astrophysically motivated parameter space. Furthermore, at the HL‑LHC luminosity of $3\ \mathrm{ab}^{-1}$, within the region defined by $100 \lesssim m_A \lesssim 220$ GeV and $m_{h^{\pm}} \lesssim 300$ GeV, the hadronic signal consistently achieves a significance that exceeds $2\sigma$. Consequently, the detectable parameter space extends to $30 \lesssim \Delta^0 \lesssim 150$ GeV and $70 \lesssim \Delta^{\pm} \lesssim 230$ GeV.

In summary, interpreting these astrophysical anomalies through DM offers a valuable clue for physics beyond the Standard Model. However, relying on relic density estimations and direct detection constraints leaves critical phenomenological ambiguities. Our findings highlight the importance of cross‑scale and multi‑messenger synergies. Combining the kinematic features extracted from high‑energy colliders with deep‑underground direct searches and space‑based cosmic‑ray observations is essential. The search strategies for mono-$W/Z$ channels established here provide a robust and indispensable tool for testing the IDM explanation of these two astrophysical excesses.

\vspace{4pt}
\section*{Acknowledgment}
\noindent The authors gratefully acknowledge the valuable discussions and insights provided by the members of the China Collaboration of Precision Testing and New Physics. 
This work was supported in part by the National Natural Science Foundation of China (NNSFC) under grant No.~12335005,~12575118 and the Special funds for postdoctoral overseas recruitment, Ministry of Education of China.

\bibliographystyle{apsrev4-1}
\bibliography{references}

\end{document}